\documentclass[10pt,letter]{article}

\usepackage{amsmath,amstext,amsgen,latexsym}
\usepackage{amstext,amssymb,amsfonts,latexsym}
\usepackage{theorem}
\usepackage{pifont}
\usepackage[dvips]{graphics,epsfig}

 \newcommand{\bs}{\bigskip}
 \newcommand{\ms}{\medskip}
 \newcommand{\n}{\noindent}
 
 \newcommand{\hs}[1]{\hspace*{ #1 mm}}
 \newcommand{\vs}[1]{\vspace*{ #1 mm}}

 \newcommand{\setempty}{\mathrm{\O}}
 \newcommand{\real}{\mathbb{R}}
 
 \newcommand{\nat}{\mathbb{N}}
 \newcommand{\integer}{\mathbb{Z}}
 \newcommand{\rational}{\mathbb{Q}}
 \newcommand{\complex}{\mathbb{C}}

 \newcommand{\dom}{\mbox{dom}}

 \newcommand{\co}{\mathrm{co}\mbox{-}}

 \newcommand{\ie}{\textrm{i.e.},\hspace*{2mm}}
 \newcommand{\eg}{\textrm{e.g.},\hspace*{2mm}}

 \newcommand{\CC}{{\cal C}}
 \newcommand{\FF}{{\cal F}}
 \newcommand{\DD}{{\cal D}}
 \newcommand{\GG}{{\cal G}}
 \newcommand{\TT}{{\cal T}}

 \newcommand{\dl}{\mathrm{L}}

 \newcommand{\np}{\mathrm{NP}}

 \newcommand{\cequalp}{\mathrm{C}_{=}\mathrm{P}}

 \newcommand{\nqp}{\mathrm{NQP}}
 
 \newcommand{\synplin}{1\mbox{-}\mathrm{synPLIN}}
 \newcommand{\syncequallin}{1\mbox{-}\mathrm{synC}_{=}\mathrm{LIN}}
 \newcommand{\reg}{\mathrm{REG}}
 
 \newcommand{\cfl}{\mathrm{CFL}}

 \newcommand{\dlin}{1\mbox{-}\mathrm{DLIN}}
 \newcommand{\plin}{1\mbox{-}\mathrm{PLIN}}
 
 \newcommand{\rdlin}{1\mbox{-}\mathrm{revDLIN}}
 \newcommand{\bplin}{1\mbox{-}\mathrm{BPLIN}}
 \newcommand{\nlin}{1\mbox{-}\mathrm{NLIN}}
 \newcommand{\alin}{1\mbox{-}\mathrm{ALIN}}
 \newcommand{\cequallin}{1\mbox{-}\mathrm{C}_{=}\mathrm{LIN}}
 
 \newcommand{\splin}{1\mbox{-}\mathrm{SPLIN}}
 \newcommand{\paritylin}{1\mbox{-}\!\oplus\!\mathrm{LIN}}
 \newcommand{\modlin}[2]{1\mbox{-}\mathrm{MOD_{#1,#2}LIN}}
 \newcommand{\sigmalinh}[1]{1\mbox{-}\Sigma^{\mathrm{LIN}}_{ #1 }}
 \newcommand{\pilinh}[1]{1\mbox{-}\Pi^{\mathrm{LIN}}_{ #1 }}
 \newcommand{\deltalinh}[1]{1\mbox{-}\Delta^{\mathrm{LIN}}_{ #1 }}

 \newcommand{\sharplin}{1\mbox{-}\#\mathrm{LIN}}
 \newcommand{\gaplin}{1\mbox{-}\mathrm{GapLIN}}
 \newcommand{\flin}{1\mbox{-}\mathrm{FLIN}}

 \newcommand{\nqlin}{1\mbox{-}\mathrm{NQLIN}}
 \newcommand{\bqlin}{1\mbox{-}\mathrm{BQLIN}}
 
 \newcommand{\nlinmv}{1\mbox{-}\mathrm{NLINMV}}
 \newcommand{\nlinsv}{1\mbox{-}\mathrm{NLINSV}}

 \newcommand{\sharpp}{\#\mathrm{P}}
 
 \newcommand{\lpf}{\mathrm{LPF}}

 \newcommand{\track}[2]{[{\tiny \begin{array}{c} #1 \\%
      #2 \end{array} }]}

 \newcommand{\IFF}{\Longleftrightarrow}

 \def\bbox{\vrule height6pt width6pt depth1pt}
 \theoremstyle{plain}
 \theoremheaderfont{\bfseries}
 \newtheorem{theorem}{Theorem}[section]
 \newtheorem{lemma}[theorem]{Lemma}
 \newtheorem{proposition}[theorem]{Proposition}
 \newtheorem{corollary}[theorem]{Corollary}

 {\theorembodyfont{\rmfamily}
     }
 {\theorembodyfont{\rmfamily} }
 {\theorembodyfont{\rmfamily} }

 \newenvironment{proof}{\par \noindent
            {\bf Proof. \hs{2}}}{\hfill$\Box$ \vspace*{3mm}}
 
 \newenvironment{proofof}[1]{\vspace*{5mm} \par \noindent
         {\bf Proof of #1.\hs{2}}}{\hfill$\Box$ \vspace*{3mm}}

 \newcommand{\ceilings}[1]{\lceil #1 \rceil}

 \newcommand{\Supp}[2]{{\mathrm{Supp}_#1}(#2)}

 \newcommand{\gslrat}{\mathrm{GSL}_{rat}}
 \newcommand{\slrat}{\mathrm{SL}_{rat}}
 \newcommand{\gslrateq}{\mathrm{GSL}^=_{rat}}
 \newcommand{\slrateq}{\mathrm{SL}^=_{rat}}
 \newcommand{\gafrat}{\mathrm{1GAF}_{rat}}

 \newcommand{\dtime}[1]{\mathrm{DTime}(#1)}
 \newcommand{\rdtime}[1]{\mathrm{revDTime}(#1)}
 \newcommand{\ntime}[1]{\mathrm{NTime}(#1)}
 \newcommand{\bptime}[1]{\mathrm{BPTime}(#1)}

 \newcommand{\sigmatime}[2]{\Sigma_{#1}\mathrm{Time}(#2)}
 \newcommand{\pitime}[2]{\Pi_{#1}\mathrm{Time}(#2)}

 \newcommand{\ignore}[1]{}
 \newcommand{\low}{\mathrm{low}}

 \newcommand{\ket}[1]{| #1 \rangle}

 \newcommand{\cent}{{|}\!\!\mathrm{c}} 

 \newcommand{\onesharplin}{1\mbox{-}\#\mathrm{LIN}}
 \newcommand{\onecequallin}{1\mbox{-}\mathrm{C}_{=}\mathrm{LIN}}

 \newcommand{\oneplin}{1\mbox{-}\mathrm{PLIN}}
 \newcommand{\onenlin}{1\mbox{-}\mathrm{NLIN}}

\begin{document}
\pagestyle{plain}
\setcounter{page}{1}
\begin{center}
{\Large {\bf Theory of One Tape Linear Time Turing Machines}}
\footnote{An earlier version appeared in the 
Proceedings of the 30th SOFSEM Conference on Current 
Trends in Theory and Practice of Computer Science, 
Lecture Notes in Computer Science, 
Vol.2932, pp.335--348, Springer-Verlag,
January 24--30, 2004. This work was in part supported by
the Natural Sciences and Engineering Council of Canada.} \ms\\
\begin{center}
{\sc Kohtaro Tadaki}$^1$\footnote{Present address:
21st Century COE Program:
Research on Security and Reliability in 
Electronic Society,
Chuo University,
1-13-27 Kasuga, Bunkyo-ku, Tokyo 112-8551, Japan.
This work was partly done while he was visiting 
the University of Ottawa between 
November 1 and December 1 in 2001.}\hspace{1cm}
{\sc Tomoyuki Yamakami}$^2$\footnote{Present address:
School of Computer Science and Engineering, University of Aizu,
90 Kami-Iawase, Tsuruga, Ikki-machi, Aizu-Wakamatsu,
Fukushima 965-8580, Japan.}\hspace{1cm}
{\sc Jack C. H. Lin}$^2$
\end{center}
\ms

${}^1$
ERATO Quantum Computation and Information Project \\ 
Japan Science and Technology Corporation, 
Tokyo, 113-0033 Japan \\

${}^2$
School of Information Technology and Engineering \\
University of Ottawa, 
Ottawa, Ontario, Canada K1N 6N5 
\end{center}

\paragraph{Abstract.}
A theory of one-tape (one-head) linear-time Turing machines is 
essentially different from
its polynomial-time counterpart since
these machines are 
closely related to finite state automata. 
This paper discusses structural-complexity 
issues of one-tape Turing
machines of various types (deterministic, 
nondeterministic,
reversible, alternating, probabilistic, 
counting, and quantum Turing machines) 
that halt in linear time, where the running
time of a machine is defined as the length 
of any longest computation path. We explore structural properties 
of one-tape linear-time Turing machines and
clarify how the machines' resources affect
their computational patterns and power.

\bs

\n\textbf{Key words.}
one-tape Turing machine, crossing sequence, finite state automaton,
regular language, one-way function, low set, advice, many-one reducibility

\section{Prologue}

Computer science has revolved around the study of computation incorporated with the analysis and development of fast and efficient algorithms. The notion of a {\em Turing machine}, proposed by Turing \cite{Tur36,Tur37}
and independently by Post \cite{Pos36} in the mid 1930s, is now
regarded as a mathematical model of many existing 
computers. This machine model has long 
been a foundation of extensive studies in computational complexity
theory. Early research unearthed the significance of various
restrictions on the {\em resources} of machines: for instance, the
number of work tapes, the number of heads, execution time bounds, memory
space bounds, and machine types in use.  This paper aims at the better understanding of how various resource restrictions directly affect the patterns and the power of computations.

The number of work tapes and also machine types of time-bounded Turing machines significantly alter their computational power. For instance, two-tape Turing machines are shown to be more powerful than any one-tape Turing machines \cite{DG84,Rab63}. Even on the model of multiple-tape Turing
machines, Paul, Pippenger, Szemeredi, and Trotter
\cite{PPST83} proved in the early 1980s that linear-time
nondeterministic Turing machines are more powerful than their
deterministic counterparts. 

Of particular interest in this paper is the model of one-tape (or single-tape) one-head linear-time Turing machines, 
apart from well-studied polynomial-time machines. Not surprisingly, this rather simple model proves a close tie to finite state automata. Despite its simplicity, such a model still offers complex structures. As a result, 
a theory of one-tape (one-head) linear-time 
complexity draws a picture quite different from multiple-tape models as well as polynomial-time models. It is thus possible for us to prove, for instance, the collapses and separations of numerous one-tape linear-time complexity classes without any unproven assumption, such as the existence of one-way functions.

Hennie \cite{Hen65} made the first major contribution to the theory of one-tape linear-time Turing machines in the mid 1960s. He demonstrated that no one-tape
linear-time deterministic Turing machine can be more powerful than 
deterministic finite state automata. To prove his result, Hennie
described the behaviors of a Turing machine in terms of the sequential
changes of the machine's internal states at the time when the tape
head crosses a boundary of two adjacent tape cells. Such a sequence of
state changes is known as a {\em crossing sequence} generated  
at this boundary. Using this 
technical tool, he argued
that (i) any one-tape linear-time deterministic Turing machine has
short crossing sequences at every boundary and (ii) if any crossing
sequence of the machine is short, then this machine recognizes only a
regular language. Using the {\em non-regularity} 
measure of Dwork and Stockmeyer \cite{DS90}, the 
second claim asserts that any language accepted by a machine with short crossing sequences has constantly-bounded non-regularity. 
Extending Hennie's
argument, Kobayashi \cite{Kob85} later showed that any language recognized by 
one-tape $o(n\log n)$-time deterministic Turing machines should be 
regular as well. This time bound $o(n\log n)$ is actually optimal since certain one-tape
$O(n\log n)$-time deterministic Turing machines can recognize
non-regular languages.

Unlike polynomial-time
computation, one-tape linear-time nondeterministic
computation is sensitive to the
definition of the machine's running time. Such sensitivity is also
observed in average-case complexity theory \cite{Yam97}.  
By taking his {\em weak definition} that defines the running time of a
nondeterministic Turing machine to be the length of a 
``shortest'' accepting path, Michel \cite{Mic91} 
demonstrated that one-tape nondeterministic Turing machines 
running in linear time (in the sense of his weak definition) 
solve even $\np$-complete problems. Clearly, his weak definition
gives an enormous power to one-tape nondeterministic machines and therefore it does not seem to offer any interesting features of time-bounded  nondeterminism. On the contrary, the {\em strong definition} (in Michel's term) requires the running time to be the length of any ``longest'' (both accepting and rejecting) computation path.
This strong definition provides us with a reasonable basis to study the effect of linear-time bounded computations. We therefore adopt his 
strong definition of running time and, throughout this paper, 
all one-tape time-bounded Turing machines are assumed to 
accommodate this strong definition. 
By expanding Kobayashi's result, we
prove that one-tape $o(n\log n)$-time nondeterministic Turing machines recognize only regular languages. 

The model of {\em alternating Turing machines} of Chandra, Kozen, and 
Stockmeyer \cite{CKS81} naturally expand the model of nondeterministic machines. The number of alternations of such an alternating Turing machine seems to enhance the computational power of the machine; however, our strong definition of running time makes it possible for us to prove that a constant number of alternations do not give any additional computational power to  one-tape linear-time alternating Turing machines; namely, such machines recognize only regular languages. 

Apart from nondeterminism, {\em probabilistic Turing machines} 
with fair coin tosses of Gill \cite{Gil77}, can present distinctive features.  
Any language recognized by a certain one-head one-way
probabilistic finite automaton with unbounded-error probability 
is known as a {\em stochastic} language \cite{Rab63}.
By employing 
a crossing sequence argument, we can 
show that any language recognized
by one-tape linear-time probabilistic Turing machines 
with unbounded-error probability is just stochastic. 
This collapse result again proves a close
relationship between one-tape linear-time Turing machines and finite
state automata.  

The model of Turing machines, nonetheless, presents distinguishing looks when we discuss functions rather than languages. Beyond the framework of formal language theory, Turing machines are capable 
of computing (partial multi-valued) 
functions by simply modifying their tape contents and producing output strings (which are sometimes viewed as numbers). 
Such functions also serve as {\em many-one reductions} between two languages. 
To explore the structure of language classes, we introduce various
types of ``many-one one-tape linear-time'' reductions.  
Nondeterministic many-one reducibility, for instance, plays an
important role in showing the aforementioned collapse of alternating 
linear-time complexity classes. Naturally, we can view many-one reducibility as oracle mechanism of the simplest form. In terms of such oracle computation, we can easily prove the existence of an oracle that separates the one-tape linear-time nondeterministic complexity class from its deterministic counterpart. 

The existence of a {\em one-way function} is a key to the building of secure cryptosystems. Intuitively, a one-way function is a function that is easy to compute but hard to invert. Restricted to one-tape linear-time deterministic computation, we can show that no one-way function exists. 

The number of accepting computation paths of a time-bounded nondeterministic Turing machine has been a crucial player in computational complexity theory.  With the notion of {\em counting Turing machines}, Valiant \cite{Val79} initiated a systematic study in the late 1970s on the structural properties of counting such numbers. Counting Turing machines have been since then used to study the complexity of ``counting'' on numerous issues in computer science. The functions computed by these machines are called {\em counting functions} and complexity classes of languages defined in terms of such counting functions are generally referred to as {\em counting classes}. We show that counting functions computable by one-tape linear-time counting Turing machines are more powerful than deterministically computable functions. By contrast, we also prove that certain counting classes induced from one-tape linear-time counting Turing machines collapse to the family of regular languages.

The latest variant of the Turing machine model is a quantum Turing
machine, which is seen as an extension of a probabilistic Turing
machine. While a probabilistic Turing machine is based on classical physics, a quantum Turing machine is based on quantum physics.
The notion of such machinery was introduced by
Deutsch~\cite{Deu85} and later 
reformulated by Bernstein and Vazirani \cite{BV97}. 
Of all the known types of quantum Turing machines, we study only
the following two machine types: bounded-error quantum Turing machines \cite{BV97} and ``nondeterministic'' quantum
Turing machines \cite{ADH97}. We give a characterization 
of one-tape linear-time ``nondeterministic'' quantum Turing 
machines in terms of counting Turing machines.

We also discuss supplemental mechanism called {\em advice} to enhance the computational power of Turing machines. Karp and Lipton \cite{KL82} formalized the notion of advice, which means additional information supplied to underlying computation besides an original input. We adapt their notion  in our setting of one-tape Turing machines as well as finite state automata. We can demonstrate the existence of context-free languages that cannot be recognized by any one-tape linear-time deterministic Turing machines with advice.   

\section{Fundamental Models of Computation}\label{sec:model}

This paper uses a standard definition of a
Turing machine (see, \eg \cite{DK00,HU79}) as a computational model. 
Of special interest are one-tape one-head Turing machines of various machine types. 
Here, we give brief descriptions of  
fundamental notions and notation associated 
with our computational model. 

Let $\integer$, $\rational$, $\real$ be the sets 
of all integers, of all rational numbers, of all 
real numbers, respectively. 
In particular, let $\real^{\geq0}$ be 
$\{r\in\real\mid r\geq0\}$. Moreover, let 
$\nat$ denote the set of all 
natural numbers (\ie non-negative integers)
and set $\nat^+=\nat-\{0\}$. For any two integers $n,m$ with $n\leq m$, an integer interval $[n,m]_{\integer}$ means the set $\{n,n+1,n+2,\ldots,m\}$. We assume that all logarithms are to the base two. Throughout this paper, we use the 
notation $\Sigma$ ($\Sigma_1$, $\Sigma_2$, etc.) to denote an
arbitrary nonempty finite alphabet. A {\em string} over alphabet 
$\Sigma$ is a finite sequence of elements from $\Sigma$ and 
$\Sigma^*$ denotes
the collection of all finite strings over $\Sigma$.  
Note that the {\em empty string} 
over any alphabet is always denoted $\lambda$. Let $\Sigma^{+} = \Sigma^* - \{\lambda\}$. 
For any string $x$ in $\Sigma^*$,
$|x|$ denotes the {\em length} of $x$ (\ie the 
number of symbols in $x$). A {\em language} (or simply a ``set'') 
over alphabet $\Sigma$ is
a subset of $\Sigma^*$,
and a {\em complexity class} is a collection of 
certain languages.
The {\em complement} of $A$ is the difference $\Sigma^*-A$,
and it is often denoted $\overline{A}$ if $\Sigma$ is 
clear from the context.
For any complexity class $\CC$, the {\em complement} 
of $\CC$, denoted $\co\CC$, is the collection of all languages whose
complements belong to $\CC$. 

We often use {\em multi-valued partial functions} as well as single-valued total functions. For any multi-valued partial function $f$ mapping from a set $D$ to another set $E$, 
$\dom(f)$ denotes the {\em domain} of $f$, namely, $\dom(f)=\{x\in D \mid f(x)\text{ is defined}\}$ and, for each $x\in\dom(f)$, $f(x)$ is a subset of $E$. Whenever $f$ is single-valued, we write ``$f(x)=y$'' instead of ``$y\in f(x)$'' by identifying the set $\{y\}$ with $y$ itself. 
Notice that {\em total} functions are also partial functions. The {\em characteristic function} 
$\chi_A$ of a language $A$ over $\Sigma$ is defined as, for any string $x$ in $\Sigma^*$, $\chi_A(x)=1$ 
if $x\in A$ and $\chi_A(x)=0$ otherwise. For any single-valued total function $g$ from $\nat$ to $\nat$, $O(g(n))$ denotes the set of all single-valued total functions $f$ such that $f(n)\leq c\cdot g(n)$ for all but finitely many numbers $n$ in $\nat$, where $c$ is a positive constant independent of $n$. Similarly, $o(g(n))$ is the set of all functions $f$ such that, for every positive constant $c$, $f(n)< c\cdot g(n)$ for all but finitely many numbers $n$ in $\nat$.  

Let us give the basic definition of one-tape (one-head) 
Turing machines. A {\em one-tape (one-head) Turing machine} 
(abbreviated 1TM) is a septuple
$M=(Q,\Sigma,\Gamma,\delta,q_0,q_{acc},q_{rej})$,
where $Q$ is a finite set of (internal) states,
$\Sigma$ is a nonempty finite input 
alphabet, 
$\Gamma$ is a finite tape alphabet including $\Sigma$,
$q_0$ in $Q$ is an initial state,
$q_{acc}$ and $q_{rej}$ in $Q$
are an accepting state and a rejecting state, 
respectively,
and $\delta$ is a transition function.
In later sections, we will define different types of transition 
functions $\delta$, which give rise to various types of 1TMs.
A {\em halting state} is either $q_{acc}$ or $q_{rej}$.
Our 1TM is equipped only with one input/work tape such that
(i) the tape stretches infinitely to both ends,
(ii) the tape is sectioned by cells,
and (iii) all cells in the tape are indexed with 
integers.
The tape head starts at the cell indexed $0$ 
(called the {\em start cell})
and either moves to the right (R), 
moves to the left (L), or stays still (N).

A {\em configuration} of a 1TM $M$, which represents 
a snapshot of a ``computation,'' is a triplet of an internal state, 
a head position, and a tape content of $M$. The initial configuration of $M$ on input $x$ is the configuration in which $M$ is in internal state $q_0$ with the head scanning the start cell 
and the string $x$ is written in an 
input/work tape, surrounded by the blank symbols, in such a way that the leftmost symbol of $x$ is in the start cell.
A computation of a 1TM $M$
generally forms a tree (called a 
{\em computation
tree}) whose nodes are certain configurations of $M$. The root of such a computation tree is an initial configuration, leaves are final configurations, and every non-root node is obtained from its parent node by a single application of $\delta$. Each path of a computation tree, from its root to a certain leaf is referred to as a {\em computation path}. An {\em accepting} 
(a {\em rejecting}, a {\em halting}, resp.) {\em computation path} is
a path terminating in an accepting (a rejecting, a halting, resp.) configuration.
We say that a TM 
{\em halts} on input $x$
if {\em every} computation path of $M$ on input $x$ eventually
reaches a certain halting state.  
Of particular importance is the
synchronous notion for 1TMs.
A 1TM is said to be {\em synchronous}
if all computation paths terminate at the same time on 
each input; namely, all the computation paths have the same length. 

Throughout this paper, we use the term ``{\em running time}'' 
for a 1TM $M$ taking input $x$, denoted $\mathrm{Time}_{M}(x)$, 
to mean the 
height of the computation tree produced by the execution of
$M$ on input $x$; in other words,
the length of any longest computation path (no matter what halting state the machine reaches) of $M$ on $x$.
We often use the notation $T(n)$ to denote a time-bounding function of a 
given 1TM that
maps $\nat$ to $\nat$. Furthermore, 
a ``linear function'' means a function of the 
form $cx+d$ for a
certain constant $c,d\in\real^{\geq0}$.
A 1TM $M$ is said to run in {\em linear time}
if its running time $\mathrm{Time}_{M}(x)$ on any input $x$
is upper-bounded by $f(|x|)$ for a certain linear function $f$.

Although our machine has only one input/work tape,
the tape can be split into a constant number of {\em tracks}.
To describe such tracks,
we use the following notation. For any pair of 
symbols $a,b\in\Sigma$,
$\track{a}{b}$ denotes the special tape symbol 
for which $a$ is written
in the upper track and $b$ is written in the 
lower track of the same cell. By extending this notion,
for any strings $x,y\in\Sigma^*$ with $|x|=|y|$, 
we write $\track{x}{y}$ to denote
the concatenation $\track{x_1}{y_1}\track{x_2}{y_2}\cdots
\track{x_n}{y_n}$ if $x=x_1x_2\cdots x_n$ and 
$y=y_1y_2\cdots y_n$, where all $x_i$'s and $y_i$'s are in $\Sigma$.

For the definition of language recognition, we need to impose 
certain reasonable {\em accepting criteria} as well as  
{\em rejecting criteria} onto our 1TMs   
to define the set of ``accepted'' input strings. 
With such criteria, we say that a 1TM {\em recognizes} a language $A$ 
if, for every string $x$, (i) if $x\in A$ then $M$  
halts on input $x$ and satisfies the accepting criteria and (ii)
if $x\not\in A$ then $M$ halts and satisfies the rejecting criteria. 

The non-regularity measure has played a key role in automata theory.
For any pair $x$ and $y$ of strings and any integer 
$n\in\nat$,
we say that $x$ and $y$ are {\em $n$-dissimilar} with 
respect to a given language $L$
if there exists a string $z$
such that (i) $|xz|\leq n$ and $|yz|\leq n$ and 
(ii) $xz\in L\;\IFF\;
yz\not\in L$.
For each $n\in\nat$, define $N_{L}(n)$
(the {\em non-regularity} measure of $L$ at $n$) to be the 
maximal cardinality
of a set in which any distinct pair is $n$-dissimilar 
with respect to $L$
\cite{DS90}.
It is immediate from the Myhill-Nerode theorem \cite{HU79}
that a language $L$ is regular if and only if $N_{L}(n)=O(1)$ \cite{DS90}.
This is further improved by the results of Karp \cite{Kar67} 
and of Ka\c{n}eps and Freivalds \cite{KF90} as follows:
a language $L$ is regular if and only if $N_{L}(n)\leq \frac{n}{2}+1$ 
for all but
finitely-many numbers $n$ in $\nat$.

We assume the reader's familiarity with the notion of 
{\em finite (state) automata}
(see, \eg \cite{HU69,HU79}). 
The class of all {\em regular} languages is denoted $\reg$, where
a language is called {\em regular} if it is recognized by a certain 
(one-head one-way) deterministic finite automaton. The languages recognized by (one-head one-way) nondeterministic push-down automata are called 
{\em context-free} and the notation $\cfl$ denotes the collection of all context-free languages. 

A {\em rational (one-head) one-way 
generalized probabilistic finite automaton}
(for short, {\em rational 1GPFA}) \cite{Tur68,Tur69b} is a quintuple
$N=(Q, \Sigma,\pi,\{T(\sigma)\,|\,\sigma\in\Sigma\},\eta)$,
where (i) $Q$ is a finite set of states,
(ii) $\Sigma$ is a finite alphabet,
(iii) $\pi$ is a row vector of length $|Q|$ having rational 
components,
(iv) for each $\sigma\in\Sigma$,
$T(\sigma)$ is an $|Q|\times|Q|$ matrix 
whose elements are rational numbers,
and (v) $\eta$ is a column vector of $|Q|$ rational entries.
A {\em word matrix} $T(x)$ of $N$ on input string 
$x\in\Sigma^*$  is defined as
$T(\lambda) = I$ for the empty string $\lambda$, 
where $I$ is the identity matrix of order
$|Q|$, and $T(x_1\dots x_k) = T(x_1)\dots
T(x_k)$ for $x_1,\dots,x_k\in\Sigma$.
For each $x\in\Sigma^*$, the {\em acceptance function} 
$p_N(x)$ is defined to be $\pi\,T(x)\,\eta$.
A matrix $T$ is called {\em stochastic} if every row of 
$T$ sums up to exactly $1$. 
A {\em rational (one-head) one-way 
probabilistic 
finite automaton}
(for short, {\em rational 1PFA}) \cite{Rab63} $N$ is a rational 1GPFA
$(Q, \Sigma,\pi,\{T(\sigma)\,|\,\sigma\in\Sigma\}, \eta)$
such that (i) $\pi$ is a stochastic row vector whose entries are all
nonnegative, (ii) for each symbol $\sigma\in\Sigma$,
$T(\sigma)$ is stochastic with nonnegative components, and (iii)
$\eta$ is a column vector whose components are either $0$ or $1$.
{}From this $\eta$, we define the set $F$ of all final states of $N$ as $F =\{a\in Q\mid \text{the $a$th entry of $\eta$ is 1}\}$.
Moreover,  since
$p_N(x)$ equals the probability of $N$ accepting $x$,
$p_{N}(x)$ is called the 
{\em acceptance probability} of $N$ on the input $x$.

Let $\varepsilon$ be any rational number.
For each rational 1GPFA $N$, let 
$L(N,\varepsilon) = \{\,x\in\Sigma^*\mid p_N(x)>\varepsilon\}$
and $L^{=}(N,\varepsilon) = 
\{\,x\in\Sigma^*\mid p_N(x)=\varepsilon\}$, where 
$\varepsilon$ is called a {\em cut point} of $N$.
Let $\gslrat$ and $\slrat$ denote the collections of all sets 
$L(N,\varepsilon)$ for certain rational 1GPFAs $N$ and for 
certain rational 1PFAs, respectively, where $\varepsilon$
is a certain rational number. Similarly, 
$\gslrateq$ and $\slrateq$ are defined from $\gslrat$ and $\slrat$, respectively, by substituting 
$L^{=}(N,\varepsilon)$ for $L(N,\varepsilon)$. 
Sets in $\slrat$ are known as {\em stochastic} languages \cite{Rab63}.
Turakainen \cite{Tur69b} demonstrated the equivalence
of $\gslrat$ and $\slrat$. With a similar idea, we can 
show that $\gslrateq=\slrateq$. The proof of this 
claim is left to the avid reader.

\section{Deterministic and Reversible Computations}
\label{sec:deterministic}

Of all computations, 
deterministic computation is one of the most intuitive 
types of computations. We begin this section 
with reviewing the major results of Hennie 
\cite{Hen65} and Kobayashi \cite{Kob85} on one-tape 
deterministic Turing machines.
A {\em deterministic 1TM}, embodying a sequential computation,
is formally defined by a transition function $\delta$ that maps 
$(Q-\{q_{acc},q_{rej}\})\times \Gamma$ to 
$Q\times \Gamma\times \{L,N,R\}$. Since the
notation $\mathrm{DLIN}$ is widely used for the model of 
multiple-tape linear-time Turing machines, we rather use the 
following new notations to emphasize our model 
of one-tape Turing machines. The
general notation $1\mbox{-}\dtime{T(n)}$ denotes the collection 
of all languages
recognized by deterministic 1TMs running in $T(n)$ time. 
Given a set ${\cal T}$
of time-bounding functions, $1\mbox{-}\dtime{{\cal T}}$ 
stands for the union
of $1\mbox{-}\dtime{T(n)}$'s over all functions $T$ in 
${\cal T}$. The {\em one-tape deterministic linear-time} 
complexity class
$\dlin$ is then defined to be $1\mbox{-}\dtime{O(n)}$.

Earlier, Hennie \cite{Hen65} proved that
$\reg=\dlin$ by employing a so-called 
{\em crossing sequence argument}. 
Elaborating Hennie's argument, Kobayashi
\cite{Kob85} substantially improved Hennie's result by showing
$\reg=1\mbox{-}\dtime{o(n\log n)}$.  
This time-bound $o(n\log n)$ is
optimal because $1\mbox{-}\dtime{O(n\log n)}$ 
contains certain non-regular
languages, \eg $\{a^nb^n\mid n\in\nat\}$ 
and $\{a^{2^n}\mid
n\in\nat\}$. These facts establish the fundamental collapse and separation results
concerning deterministic 1TMs.
 
\begin{proposition}\label{prop:Hennie}\hs{-2}{\rm \cite{Hen65,Kob85}}
$\reg= 1\mbox{-}\dtime{o(n\log n)}
\subsetneqq 1\mbox{-}\dtime{O(n\log n)}$.
\end{proposition}

In the early 1970s, Bennett \cite{Ben73} initiated 
a study of
reversible computation. Reversible computations
have recently drawn wide attention from physicists as well as
computer scientists in
connection to quantum computations. We adopt 
the following definition of a (deterministic) 
reversible Turing machine
given by Bernstein and Vazirani \cite{BV97}. 
A {\em (deterministic) reversible 1TM} is a 
deterministic 1TM of
which each configuration has at most one 
predecessor configuration.
We use the notation
$1\mbox{-}\rdtime{T(n)}$ to denote the 
collection of all languages 
recognized by $T(n)$-time reversible 1TMs and 
define $1\mbox{-}\rdtime{{\cal T}}$ to be 
$\bigcup_{T\in\TT}1\mbox{-}\rdtime{T(n)}$. Finally, let
$\rdlin = 1\mbox{-}\rdtime{O(n)}$. Obviously, 
$\rdlin$ is a subset of $\dlin$.

Kondacs and Watrous \cite{KW97}
demonstrated that any one-head one-way 
deterministic finite automaton can be
simulated in linear time 
by a certain one-head two-way deterministic reversible finite 
automaton.  Since any
one-head two-way deterministic reversible finite automaton is 
indeed a reversible 1TM, we 
obtain that $\reg\subseteq \rdlin$.  
Proposition \ref{prop:Hennie} thus concludes:

\begin{proposition}\label{trev}
  $\reg = \rdlin = 1\mbox{-}\rdtime{o(n\log n)}$.
\end{proposition}

The computational power of a Turing machine can be enhanced by 
supplemental information given besides inputs.
Karp and Lipton \cite{KL82} introduced the notion of such extra information under the name of {\em advice}, which is given depending only on the size of input. Damm and Holzer \cite{DH95} later considered finite automata that take the Karp-Lipton type advice. To make most of 
the power of advice, 
we should take a slightly different formulation for our 1TMs. 
In this paper, for any complexity class $\CC$ defined in terms of Turing machines (including finite state automata as special cases), the notation $\mathrm{REG}/n$ is used to represent the
collection of all languages $A$ for which there 
exist an alphabet $\Sigma$, a deterministic finite automaton $M$ working with another alphabet, 
and a total function\footnote{As standard in 
computational complexity theory, we allow non-recursive 
advice functions in general.}
$h$ from $\nat$
to $\Sigma^*$ with $|h(n)|=n$ (called an {\em advice function}) 
satisfying that, for every
$x\in\Sigma^*$, $x\in A$ if and only if 
$\track{x}{h(|x|)}\in L(M)$.
For instance, the context-free language 
$L_{eq} = \{ 0^n1^n \mid n\in\nat \}$ belongs to $\reg/n$. More generally, every language $L$, over alphabet $\Sigma$, whose restriction $L\cap\Sigma^n$ for each length $n$ has cardinality bounded from above by a certain constant, independent of $n$, belongs to $\mathrm{REG}/n$, because the advice can encode a finite look-up table for length $n$.
  
This gives the obvious separation $\reg\subsetneqq \reg/n$.
On the contrary, $\reg/n$ cannot include $\cfl$ since, 
as we see below, the non-regular language 
$Equal=\{x\in\{0,1\}^*\mid \#_0(x)=\#_1(x)\}$, 
where $\#_i(x)$ denotes the number of occurrences of the symbol $i$ 
in $x$, is situated outside of $\reg/n$. This result 
will be used in Section \ref{sec:counting}.  

\begin{lemma}\label{non-reg-advice}
The language $Equal$ is not in $\reg/n$. 
Hence, $\cfl\nsubseteq \reg/n$.
\end{lemma}

\begin{proof}
Let $\Sigma=\{0,1\}$.
Assuming that $Equal\in\reg/n$, choose a deterministic 
finite automaton $M=(Q,\Sigma,q_0,F)$ 
and an advice function $h$ from $\nat$
to $\Sigma^*$ such that, for every string 
$x\in\Sigma^*$, $x\in Equal$ if and only if $\track{x}{h(|x|)}\in L(M)$.
Take $n=|Q|$.
For each number $k\in [0,n]_{\integer}$, 
$y_k$ denotes any string of length $n$ 
satisfying $\#_{0}(y_k) = k$.

There exist two {\em distinct} indices
$k,l\in [0,n]_{\integer}$ such that
(i) $y_kz_k,y_lz_l\in Equal$ for certain strings $z_k,z_l\in\Sigma^n$
and (ii) $M$ enters the same internal state 
after reading
$\track{y_k}{w_n}$ as well as $\track{y_l}{w_n}$, where $w_n$ is the first $n$ bits of $h(2n)$. Notice that such a pair $(k,l)$ 
indeed exists because $n+1>|Q|$. 
It follows from these conditions that $M$ 
also accepts the input $\track{y_kz_l}{h(|y_kz_l|)}$. Thus, 
$\#_0(y_kz_l) = \#_1(y_kz_l)$, which 
implies $\#_0(z_l)=n-k$.
However, since $\#_0(y_lz_l)=\#_1(y_lz_l)$, 
we obtain $\#_0(y_l)=k$. This contradicts the 
definition of $y_l$.
Therefore, $Equal$ is not in $\reg/n$. The second claim $\cfl\nsubseteq \reg/n$ follows from the fact that $Equal\in \cfl$.
\end{proof}

Up to now, we have viewed ``Turing machines'' as language recognizers (or language acceptors); however, unlike deterministic finite state automata, Turing machines are fully capable of computing partial functions. 
Since a 1TM $M$ has only one input/work tape, 
we need to designate the same input tape as the output 
tape of the machine as well. 
To specify an ``outcome'' of the machine, we adopt the following convention.
When the machine eventually halts with its output tape  
consisting only of a single block of non-blank symbols, say $s$, 
surrounded by the blank symbols, in a way that 
the leftmost symbol of $s$ is written in the start cell, we consider $s$ 
as the {\em valid outcome} of the machine.  

For notational convenience, we 
introduce the function class $\flin$ in the following fashion. 
A {\em total} function from $\Sigma_1^*$ to 
$\Sigma_2^*$ is in $\flin$ if there exists a 
deterministic 1TM $M$ satisfying that, on 
any input $x\in\Sigma_1^*$, (i) $M$ halts by entering the 
accepting state in time linear in $|x|$ and (ii) 
when $M$ halts, $M$ outputs $f(x)$ as a valid outcome.
When ``partial'' functions are concerned, we conventionally
regard the ``rejecting state'' as an invalid outcome. We thus
define $\flin\mathrm{(partial)}$ to be the collection of all partial
functions $f$ from $\Sigma_1^*$ to $\Sigma_2^*$ such that, for every $x\in\Sigma_1^*$, (i) 
if $x\in\dom(f)$ then $M$ enters an accepting state 
with outputting $f(x)$
and (ii) if $x\not\in\dom(f)$ then $M$ enters a
rejecting state (and we ignore the tape content). 

Historically, automata theory has also provided the machinery that can compute functions (see, \eg \cite{HU79} for a historical account). 
In comparison with $\flin$, we herein consider only so-called Mealy machines.  A {\em Mealy machine} $(Q,\Sigma,\Gamma,q_0,\delta,\nu)$ is a deterministic finite automaton $(Q,\Sigma,\Gamma,q_0,\delta)$, ignoring final states, together with a total function $\nu$ from $Q\times\Sigma$ to $\Gamma$ such that, on input $x=x_1x_2\cdots x_n$, it outputs $\nu(q_{0},x_{1})\nu(q_{1},x_{2})\cdots \nu(q_{n-1},x_{n})$, where $(q_0,q_1,\ldots,q_{n})$ is the sequence of states in $Q$ satisfying $\delta(q_{i-1},x_{i})=q_{i}$ for every $i\in [1,n]_{\integer}$. Note that a Mealy machine computes only length-preserving functions, where a (total) function is called {\em length-preserving} if $|f(x)|=|x|$ for any string $x$.
Consider the length-preserving function $f$ defined by
$f(x_1x_2\cdots x_n) = x_nx_1\cdots x_{n-1}$
for any $x_1,x_2,\ldots,x_n\in\{0,1\}$. It is clear that no Mealy machine can compute $f$. We therefore obtain the following proposition.

\begin{proposition}
There exists a length-preserving function in $\flin$ that
cannot be computed by any Mealy machines.
\end{proposition}

\section{Nondeterministic Computation}
\label{sec:nondeterministic}

Nondeterminism has been widely studied in the 
literature since many
problems arising naturally in computer 
science have nondeterministic
traits. In a nondeterministic computation,
a Turing machine has several choices to follow at each 
step. We expand the collapse result of deterministic 1TMs 
in Section \ref{sec:deterministic} into nondeterministic 
1TMs. We also discuss the multi-valued partial functions
computed by one-tape nondeterministic Turing machines 
and show how to simulate such functions in 
a certain deterministic manner.

\subsection{Nondeterministic Languages}\label{sec:nondet-lang}

As a language recognizer, a {\em nondeterministic 1TM} takes a 
transition function $\delta$ that maps 
$(Q-\{q_{acc},q_{rej}\})\times \Gamma$ to 
$2^{Q\times\Gamma\times\{L,N,R\}}$, where 
$2^A$ denotes the power set of $A$. An execution of a
nondeterministic 1TM produces a computation tree.
We say that a nondeterministic 
1TM $M$ {\em accepts} an input $x$ 
exactly when there exists
an accepting computation path in the 
computation tree of $M$
on input $x$.
Similar to the deterministic case, let
$1\mbox{-}\ntime{T(n)}$ denote the 
collection of all languages 
recognized by  $T(n)$-time\footnote{As stated in Section \ref{sec:model}, this paper accommodates the {\em strong definition} of running time; namely, the running time of a machine $M$ on input $x$ is the height of the computation tree produced by $M$ on $x$, independent of the outcome of the computation.} 
nondeterministic 1TMs and let
$1\mbox{-}\ntime{{\cal T}}$ be the union of all
$1\mbox{-}\ntime{T(n)}$ for all $T\in{\cal T}$. 
We define the {\em one-tape
nondeterministic linear-time} class $\nlin$ to be
$1\mbox{-}\ntime{O(n)}$.

We first expand Kobayashi's collapse result
on $1\mbox{-}\dtime{o(n\log n)}$ into $1\mbox{-}\ntime{o(n\log n)}$.

\begin{theorem}\label{ntime}
$\reg= 
1\mbox{-}\ntime{o(n\log n)}\subsetneqq 1\mbox{-}\ntime{O(n\log n)}$. 
\end{theorem}

The proof of Theorem \ref{ntime}
consists of two technical lemmas: 
Lemmas \ref{nlogn-bound} and
\ref{nbound-regular}. The first lemma has 
Kobayashi's argument in 
\cite[Theorem 3.3]{Kob85} as its core, and the second lemma
is due to Hennie \cite[Theorem 2]{Hen65}. 
For the description of the lemmas, we need to
introduce the key terminology.

Let $M$ be any type of 1TM, which is not necessarily nondeterministic.
Any boundary that separates 
two adjacent cells in
$M$'s tape is called an {\em intercell boundary}.
The {\em crossing sequence} at intercell boundary $b$
along computation path $s$ of $M$ is the sequence 
of internal states of $M$
at the time when the tape head crosses $b$,
first from left to right,
and then alternately in both directions.
To visualize the head move, let us assume that the head is
scanning tape symbol $\sigma$ at tape cell $i$ in state $p$.
An application of a transition $(q,\tau,R)\in \delta(p,\sigma)$ makes
the machine write symbol $\tau$ 
into cell $i$, enter state $q$, and then move the head 
to cell $i+1$. The state in which the machine crosses 
the intercell boundary between cell $i$ and cell $i+1$ is $q$ 
(not $p$). Similarly, if we apply a 
transition $(q,\tau,L)\in\delta(p,\sigma)$, then $q$ is the state 
in which the machine crosses the intercell boundary between 
cell $i-1$ and cell $i$.  
The {\em right-boundary} of $x$ is the 
intercell boundary
between the rightmost symbol of $x$ and its 
right-adjacent blank symbol.
Similarly, the {\em left-boundary} of $x$ is defined
as the intercell boundary between the leftmost symbol of $x$
and its left-adjacent symbol.
Any intercell boundary between the right-boundary 
and the left-boundary
of $x$ (including both ends) is called a 
{\em critical boundary} of $x$.

Lemma \ref{nlogn-bound} observes that  
Kobayashi's argument extends to nondeterministic 1TMs 
without depending on their acceptance criteria. 
For completeness, the proof of
Lemma \ref{nlogn-bound} is included in Appendix.

\begin{lemma}\label{nlogn-bound}
Assume that $T(n)=o(n\log n)$.
For any $T(n)$-time nondeterministic 1TM $M$,
there exists a constant $c\in\nat$ such that, 
for each string
$x$, any crossing sequence at any critical  
boundary in any (accepting or rejecting) 
computation path of $M$ on the input $x$ has
length at most $c$. 
\end{lemma}

In essence, Hennie \cite{Hen65} proved 
that any deterministic
computation with short crossing sequences has constantly-bounded
non-regularity. 
We generalize his result to the nondeterministic case as
in the following lemma. Different 
from the previous lemma, Lemma
\ref{nbound-regular} relies on the 
acceptance criteria of
nondeterministic 1TMs. Nonetheless, 
Lemma \ref{nbound-regular} does not
refer to rejecting computation paths. 
For readability, the proof of Lemma 
\ref{nbound-regular} is also
placed in Appendix.

\begin{lemma}\label{nbound-regular}
Let $L$ be any language and let $M$ be any
nondeterministic 1TM that recognizes $L$. 
For each $n\in \nat$, let
$S_n$ be the set of all crossing sequences 
at any critical-boundary
along any accepting computation path of $M$ on any input
of length $\leq n$. Then, 
$N_{L}(n)\leq 2^{|S_n|}$ for all $n\in\nat$, 
where $|S_n|$ denotes the cardinality of $S_n$.
\end{lemma}

Since $\reg$ is closed under complementation, so is
$1\mbox{-}\ntime{o(n\log n)}$ by Theorem \ref{ntime}.
In contrast, a simple
crossing-sequence argument proves that
$1\mbox{-}\ntime{O(n\log n)}$ does not 
contain the set of all palindromes, 
$Pal=\{x\in\{0,1\}^*\mid x=x^{R}\}$, 
where $x^{R}$ is the {\em reverse} of $x$.
Since $\overline{Pal}\in 1\mbox{-}\ntime{O(n\log n)}$, 
$1\mbox{-}\ntime{O(n\log n)}$ is different 
from $\co1\mbox{-}\ntime{O(n\log n)}$.

\begin{corollary}
\sloppy The class
$1\mbox{-}\ntime{o(n\log n)}$ is closed 
under complementation, whereas 
$1\mbox{-}\ntime{O(n\log n)}$ is not 
closed under complementation. 
\end{corollary}

Reducibility between two languages has played a central role in 
the theory of NP-completeness as a measuring tool for the complexity of languages. We can see reducibility as 
a basis of ``relativization'' with oracles. For instance, Turing reducibility 
induces a typical adaptive oracle computation whereas truth-table reducibility represents a nonadaptive (or parallel) oracle computation.
Similarly, we introduce the following 
restricted reducibility into one-tape Turing machines. 
A language $A$ over alphabet $\Sigma_1$ is said to be 
{\em many-one 1-NLIN-reducible to} another language 
$B$ over alphabet $\Sigma_2$ (notationally,
$A\leq^{\nlin}_{m}B$) if there 
exist a linear function $T$ and a nondeterministic 1TM $M$ 
such that, for every string
$x$ in $\Sigma_1^*$, (i) $M$ on the input $x$ halts 
within time $T(|x|)$ with
the tape consisting only of one block of 
non-blank symbols, say $y_p$, 
on every computation path $p$, provided that 
the left-most symbol of $y_p$ must be written in the
start cell, (ii) when $M$ eventually halts, 
the tape head returns to the start cell along all 
computation paths, and (iii) $x\in A$ if and only if $y_p\in B$ 
for some {\em accepting} computation path $p$ of $M$ on the input $x$.
For any fixed set $B$, we use the notation $\nlin_{m}^{B}$ to 
denote the collection of all languages $A$ that are many-one
$\nlin$-reducible to $B$. Furthermore, for
any complexity class $\CC$, the notation 
$\nlin_{m}^{\CC}$ stands for the union of sets 
$\nlin_{m}^{B}$ over all sets $B$ in $\CC$.

A straightforward simulation shows
that $\nlin_{m}^{\reg}$ is included in
$\nlin$. More generally, we can show the following 
proposition. This result will be used in 
Section~\ref{sec:alternating}.

\begin{proposition}\label{lemma:REG-closure}
For any language $C$, $\nlin_{m}^{\nlin_{m}^{C}} 
\subseteq \nlin_{m}^{C}$. 
\end{proposition}

\begin{proof}
This proposition is essentially equivalent to the transitive property of the relation $\leq_{m}^{\onenlin}$. Let $A$, $B$, and $C$ be three arbitrary  languages and assume that $A\leq^{\nlin}_{m}B \leq^{\nlin}_{m}C$. Our goal is to show that $A\leq^{\nlin}_{m}C$. 
Take a nondeterministic 1TM $M$ that many-one $\nlin$-reduces $A$ to $B$ and another nondeterministic 1TM $M'$ that many-one $\nlin$-reduces $B$ to $C$. Now, consider the following 1TM
$N$. On input $x$, simulate $M$ on $x$, and if and when it halts with an admissible value on the tape, start
$M'$ on that value as its input. This machine $N$ is clearly nondeterministic and its running
time is $O(n)$ since so are the running times of $M$ and $M'$. 
It is not difficult to check that $N$ reduces $A$ to $C$.
\end{proof}

Similar to the many-one $\nlin$-reducibility, we can 
define the ``many-one $\dlin$-reducibility'' 
and its corresponding relativized class $\dlin_{m}^B$ for any set $B$. Although $\dlin=\nlin$, two reducibilities, many-one $\nlin$-reducibility and many-one $\dlin$-reducibility, are quite different in their power. As an example, we can construct  
a recursive set $B$ that separates between 
$\dlin_{m}^{B}$ and $\nlin_{m}^{B}$. The construction of such a set $B$ can be done by a standard diagonalization technique.

\begin{proposition}
There exists a recursive set $B$ such that 
$\dlin_{m}^{B}\subsetneqq \nlin_{m}^{B}$.
\end{proposition}

\begin{proof}
For any set $B\subseteq \{0,1\}^*$,
define $L_B=\{0^n\mid \exists x\in\{0,1\}^n[x\in B]\}$.
Obviously, $L_B$ belongs to $\nlin_m^B$ for any set $B$.

Baker, Gill, and Solovay \cite{BGS75} constructed a recursive set $B$ such that $L_B$ cannot be polynomial-time Turing reducible to $B$. In particular, $L_B$ is not many-one $\dlin$-reducible to $B$; that is, $L_B\not\in \dlin_{m}^{B}$.  
\end{proof}

\subsection{Multi-Valued Partial Functions}\label{partial-function}

Conventionally, a Turing machine that can output values is called a {\em transducer}. Nondeterministic transducers can compute multi-valued partial functions in general. Let us 
consider a nondeterministic 1TM that outputs a certain string in $\Sigma_2^*$ (whose leftmost symbol is in the start cell) along each computation path by entering a certain halting state. Similar to partial functions introduced in Section \ref{sec:deterministic}, we invalidate any rejecting computation path and let $M(x)$ denote the set of all {\em valid} outcomes of $M$ on input $x$.
In particular, when $M$ on the input $x$ enters the rejecting state along all computation paths, $M(x)$ becomes the empty set.
A multi-valued partial function $f$ from $\Sigma_1^*$ to $\Sigma_2^*$ is in $\nlinmv$ if there exists a linear-time nondeterministic 1TM $M$ such that $f(x) = M(x)$ for any string $x\in\Sigma_1^*$. Let $\nlinsv$ be the subset of $\nlinmv$,  containing only single-valued partial functions.
In contrast, $\nlinmv_{t}$ and $\nlinsv_{t}$ denote the collections of all {\em total} functions in $\nlinmv$ and in $\nlinsv$, respectively.
Clearly, $\flin\mathrm{(partial)} \subseteq\nlinsv\subsetneqq \nlinmv$ and $\flin \subseteq\nlinsv_t\subsetneqq \nlinmv_t$. 

Note that, for any function $f\in\nlinmv$, we can decide 
nondeterministically whether $x$ is in $\dom(f)$, and thus $\dom(f)$ 
belongs to the class $\nlin$, which equals $\reg$ by Theorem \ref{ntime}.

The basic relationship between functions in $\flin$ and 
languages in $\dlin$ is stated in Lemma \ref{LPF-NLINMV}. A multi-valued partial
function $f$ from $\Sigma_1^*$ to $\Sigma_2^*$ is called {\em length-preserving} if, for every $x\in\Sigma_1^*$ 
and $y\in\Sigma_2^*$, 
$y\in f(x)$ implies $|y|=|x|$.
For convenience, we write $\lpf$ to denote the collection of all length-preserving multi-valued partial functions from $\Sigma_1^*$ to $\Sigma_2^*$, where $\Sigma_1$ and $\Sigma_2$ are arbitrary nonempty finite alphabets. Moreover, for  
any multi-valued partial function $f\in\lpf$, let 
$L[f] = \{\track{x}{y}\mid y\in f(x)\}$.

\begin{lemma}\label{LPF-NLINMV}
For any multi-valued partial function $f\in\lpf$,
$f\in\nlinmv$ if and only if $L[f]$ is in $\dlin$.
\end{lemma}

\begin{proof}
Let $f$ be any length-preserving multi-valued partial function. Assume that $f$ is computed by a linear-time nondeterministic 1TM $M$. Consider the machine $N$ that behaves as follows: on input $\track{x}{y}$, nondeterministically compute $z$ in $f(x)$ from $x$ and check if $y=z$. This machine $N$ places $L[f]$ in $\nlin$, which equals $\dlin$. Conversely, assume that $L[f]$ is recognized by a linear-time nondeterministic 1TM $N$. We define another machine $M$ as follows: on input $x$, guess $y\in\Sigma^n$ (by writing $y$ in the second track), run $N$ on input $\track{x}{y}$. If $N$ accepts, output $y$. Clearly, $M$ computes $f$ and thus $f$ is in $\nlinmv$.
\end{proof}

The following major collapse result extends 
the collapse $\nlin=\reg$ shown in Section \ref{sec:nondet-lang}.

\begin{theorem}\label{nlinsv-flin}
$\nlinsv\cap \lpf = \flin\mathrm{(partial)}\cap \lpf$ 
and thus $\nlinsv_{t}\cap\lpf = \flin\cap\lpf$.
\end{theorem}

Theorem \ref{nlinsv-flin} 
is a direct consequence of the following key lemma.
We first introduce the notion of refinement.
For any two multi-valued partial functions $f$ and $g$ from $\Sigma_1^*$ to $\Sigma_2^*$, we say that $f$ is a {\em refinement} of $g$ if, for any $x\in\Sigma_1^*$, (i) $f(x)\subseteq g(x)$ (set inclusion) and (ii) $f(x)=\setempty$ implies $g(x)=\setempty$. (See, \eg Selman's paper \cite{Sel92} for this notion.)

\begin{lemma}\label{refinement}
Every length-preserving $\nlinmv$ function has a $\flin\mathrm{(partial)}$ refinement.
\end{lemma}

The crucial part of the proof of Lemma \ref{refinement} is 
the construction of a ``folding machine'' from a given 
nondeterministic 1TM. A folding machine rewrites the contents of cells in its input area, where the {\em input area} means the tape region where given input symbols are initially written. For later use, we give a general description of a folding machine.

\paragraph{Construction of a Folding Machine.} 
Let $M=(Q,\Sigma,\Gamma,\delta,q'_0,q_{acc},q_{rej})$ be any 1TM  that always halts in linear time. The {\em folding machine} $N$ is constructed from $M$ as follows. 
Choose the minimal positive integer $k$ such that $\mathrm{Time}_{M}(x)\leq k|x|$ for all inputs $x$ of length at least $3$. Notice that, since $M$'s tape head moves in both directions on its tape, $M$ can use tape cells indexed between $-2k(|x|-1)$ and $2k(|x|-1)-1$. Choose four new internal states $q_0,q_1,q_2,q_3$ not in $Q$ and introduce new internal states of the form $\track{i}{q}$ for each number $i\in [-2k,2k-1]_{\integer}$ and each internal state $q\in Q$. Let $x$ be an arbitrary string written in the input tape of $M$.

\ms

1) The machine $N$ starts in the new initial state $q_0$. If the input $x$ is empty, then $N$ immediately enters $M$'s halting state without moving its head. Hereafter, we assume that $x$ is a nonempty string of the form $\sigma_1\sigma_2\cdots \sigma_n$, where each $\sigma_i$ is a symbol in $\Sigma$. 
Note that $\sigma_1$ is written in the start cell.

\begin{figure}[t]
\begin{center}
\includegraphics*[width=12.5cm]{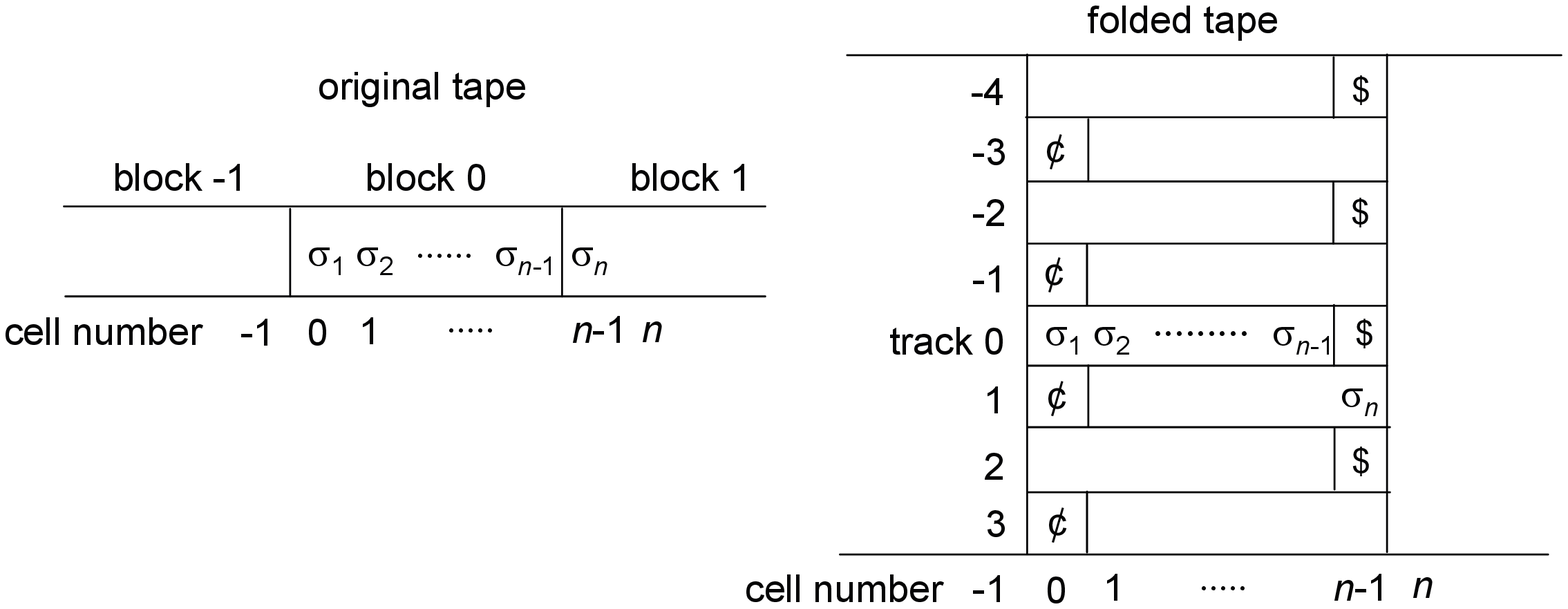}
\caption{The tape design of the folding machine $N$ where $k=2$. The original tape of $M$ is partitioned into $4k$ blocks of size $n-1$ and each block is simulated by a track in the folded tape of $N$. For instance, block 0 is simulated by track 0, and block 1 is simulated by track 1 in the reverse order.}
\label{tracks} 
\end{center}
\end{figure}

2) In this preprocessing phase, the machine $N$ re-designs its input/work tape, as shown in Figure~\ref{tracks}, by moving its head.
In the original tape of $M$, the cells indexed between $-2k(|x|-1)$ and $2k(|x|-1)-1$
are partitioned into $4k$ blocks of $|x|-1$ cells.
These blocks are indexed in order from the leftmost block to the rightmost block using integers ranging from $-2k$ to $2k-1$.
In particular, block $0$ contains the string $\sigma_1\sigma_2\cdots \sigma_{n-1}$ (without $\sigma_n$).
We split the tape of $N$ into $4k$ tracks, which are indexed from the top to the bottom using $-2k$ to $2k-1$. Intuitively, we want to simulate block $i$ of $M$'s tape using track $i$ of $N$'s folded tape. 
The machine $N$ first places the special symbol $\cent$ ({\em left end-marker}) in all tracks of odd indices and then enters the internal state $q_1$ by stepping right. The machine keeps moving its head rightward in the state $q_1$. When the head encounters the first blank symbol,
if $|x|\ge 3$ then $N$ enters the 
state $q_2$ and steps back; otherwise, $N$ enters $M$'s halting state.
In a single step, $N$ places another special symbol \$ ({\em right end-marker}) in all tracks of even indices, shifts $\sigma_n$ in track $0$ to track $1$, enters the state $q_3$, and steps to the left. The head then returns to the start cell in state $q_3$. Notice that this phase can be done in a reversible fashion.

3) The machine $N$ simulates $M$'s move by folding $M$'s tape content into $4k$ tracks of the input area.
While $M$ stays within block $i$ in state $q$, $N$ simulates $M$'s move on track $i$ with internal state $\track{i}{q}$. If $i$ is even, then $N$ moves its head in the same direction as $M$ does. Otherwise, $N$ moves the head in the opposite direction.
In particular,
at the time when $M$'s head leaves the last (first, resp.) cell of block $2j$ to its adjacent block by rewriting symbol $\sigma$ and entering the state $q$,
$N$ instead enters state $\track{2j+1}{q}$ ($\track{2j-1}{q}$, resp.),
writes symbol $\sigma$ in track $2j$ ($2j$, resp.),
and moves its head to the right (right, resp.).
On the contrary, at the time when $M$'s head leaves block $2j+1$, $M$ moves the head similarly but in the opposite direction.
It is clear that $N$'s head never visits outside of the input area. This simulation phase takes exactly the same amount of time as $M$'s.

Consider the set $S$ of all (possible) crossing sequences of the folding machine $N$. For any two crossing sequences $v,v'\in S$ and any tape symbol $\sigma$, we write $v\rightarrow_{\sigma}v'$ if $v$ is a crossing sequence of the left-boundary of $\sigma$ and $v'$ is a crossing sequence of the right-boundary of $\sigma$ along a certain computation path of $N$ on input $x\sigma y$ for certain strings $x$ and $y$. 
Along any computation path $p$ of $N$ on any nonempty input $x$, it is important to note that $v_0=()$, the {\em empty sequence}, and $v_f=(q_1,q_2)$ are respectively the unique crossing sequences at the left-boundary and the right-boundary of $x$. We can translate this computation path $p$ on input $x=\sigma_1\sigma_2\cdots\sigma_n$ into its corresponding series of crossing sequences, $v_0,v_1,\ldots,v_{n}$, satisfying the following conditions: $v_n=v_f$ and $v_{i-1}\rightarrow_{\sigma_{i}}v_{i}$ for every index 
$i\in [1,n]_{\integer}$.

Now, let us return to the proof of Lemma \ref{refinement}.

\begin{proofof}{Lemma \ref{refinement}}
Let $f$ be any length-preserving 
multi-valued partial function in $\nlinmv$. There exists a linear-time nondeterministic 1TM $M^*=(Q,\Sigma,\Gamma,\delta,q_0,q_{acc},q_{rej})$ that computes $f$. 
Consider the folding machine $N$ constructed from $M^*$. Matching the output convention of 1TMs, we need to modify this folding machine to produce the outcomes of $M^*$. 
After $N$ eventually halts, we further move the tape head leftward. 
When we reach the left end-marker, we move back the head by 
changing the current tape symbol to the symbol written in the area of 
track $0$ and track $1$ 
where the original input symbols of $N$ is written.
When we reach the right end-marker, we step right to the first blank symbol by entering a halting state (either $q_{acc}$ or $q_{rej}$) of $M^*$. Evidently,   
this modified nondeterministic 1TM produces the outcomes of $M^*$
and also enters exactly the same halting states of $M^*$. This modified machine is hereafter referred to as $N$ for our convenience.

Let $CS$ be the set of all crossing sequences of $N$. Assume that all elements in $CS$ are enumerated so that we can always find the minimal element in any subset of $CS$. 
For any two elements $v,v'\in CS$ and any symbol $\sigma$ with $v\rightarrow_{\sigma}v'$, $Symb(v,\sigma,v')$ denotes  
the output symbol written in the cell where $\sigma$ is initially written. This symbol $Symb(v,\sigma,v')$ can be easily deduced from $(v,v',\sigma)$ by tracing the tape head moves crossing a cell that initially contains the symbol $\sigma$.

Finally, we want to construct a refinement $g$ of $f$. This desired partial function $g$ is defined by a deterministic 1TM $M$ that behaves as follows. Let $n\in\nat$ and let $x=\sigma_1\sigma_2\cdots\sigma_n$ be an arbitrary input of length $n$. Set $v_0=()$ and $v_f=(q_1,q_2)$ as before.
\ms

1) In this phase, all internal states except $v_f$ are subsets of $CS$. Let $S_1=\{v_0\}$ be the initial state of $M$. Let $i\in[1,n]_{\integer}$ and  assume that $M$ currently scans the input symbol $\sigma_i$ in internal state $S_i$. We define two key sets $V_i = \{v\in S_i\mid \exists\>v'\!\in CS\;[v\rightarrow_{\sigma_i}v']\}$ and $S_{i+1} =\{v'\in CS\mid \exists\>v\in V_i\;[v\rightarrow_{\sigma_i}v']\}$. 
Intuitively, $S_{i+1}$ captures all possible nondeterministic 
moves from $S_i$. Notice that $V_i\subseteq S_i$. 
When $S_{i+1}$ is empty, $M$ enters a new rejecting state. Provided that $S_{i+1}$ is non-empty, $M$ changes the tape symbol $\sigma_i$ to $\track{\sigma_i}{V_i}$ and enters $S_{i+1}$ as an internal state by stepping to the right. Unless $x\not\in\dom(f)$, after scanning $\sigma_n$, $M$ enters the internal state $S_{n+1}$. By the property of the original folding machine, we must have $S_{n+1} = \{v_f\}$. For later convenience, let $v_n = v_f$ and $V_{n+1} = S_{n+1}$. When the tape head scans the first blank symbol, $M$ then enters the internal state $v_n$ by stepping to the left. 

2) In the beginning of this second phase, $M$ is in the state $v_n$, scanning 
the rightmost tape symbol $\track{\sigma_n}{V_n}$ in the input area. Notice that $v_n\in V_{n+1}$.  For any index $i\in[1,n]_{\integer}$, let us assume 
that $M$ scans the symbol $\track{\sigma_i}{V_i}$ in the state $v_i$, where $v_i\in V_{i+1}\subseteq S_{i+1}$. 
Since $M$ passes the first phase and enters the second phase, $V_i$ cannot be empty. Since $v_i\in S_{i+1}$, the set 
$W_i = \{v\in V_{i}\mid v \rightarrow_{\sigma_i}v_i\}$ is not empty, either. 
Choose the minimal element, say $v_{i-1}$, in $W_i$. This crossing sequence $v_{i-1}$ obviously satisfies that  $v_{i-1}\rightarrow_{\sigma_i}v_i$. 
Now, $M$ changes the symbol $\track{\sigma_i}{V_i}$ to $Symb(v_{i-1},\sigma_i,v_i)$ and moves its tape head to the left by entering $v_{i-1}$ as an internal state. After scanning $\track{\sigma_1}{V_1}$, $M$ enters the internal state $v_0$ because $V_1=S_1 =\{v_0\}$. Note that the resulting series $(v_0,v_1,v_2,\ldots,v_n)$ specifies a certain accepting computation path of $N$ and the output tape of $M$ contains the outcome produced along this particular computation path. When the tape head reaches the blank symbol, $M$ finally enters a new accepting state. This completes the description of $M$.

\ms

The above deterministic 1TM $M$ clearly produces, for each input $x$, at most one output string from the set $f(x)$. Note that, if $x\not\in\dom(f)$, all computation paths are rejecting paths, and thus $M$ never reaches any accepting state. It is therefore obvious that the partial function $g$ computed by $M$ is a refinement of $f$. 
\end{proofof}

Another application of Lemma \ref{refinement} is the non-existence of one-way functions in $\flin$. To describe the notion of one-way function in our single-tape linear-time model, we need to expand our ``track'' notation $\track{x}{y}$ to the case where $|x|$ and $|y|$ differ. To keep our notation simple, we also use the same notation $\track{x}{y}$ to express $\track{x\#^k}{y}$ if $|x|+k=|y|$ and $k\geq1$ and express $\track{x}{y\#^k}$ if $|x|=|y|+k$ and $k\geq1$, where $\#$ is a distinct ``blank'' symbol.  
A total function $f$ is called {\em one-way} if (i) $f\in\flin$ and (ii) there is no function $g\in\flin$ such that $f\left(g\left(\track{f(x)}{1^{|x|}}\right)\right) = f(x)$ for all inputs $x$. When $f$ is length-preserving, the equality $f\left(g\left({\track{f(x)}{1^{|x|}}}\right)\right) = f(x)$ can be replaced by $f(g(f(x))) = f(x)$.

\begin{proposition}
There is no one-way function in $\flin$. 
\end{proposition}

\begin{proof}
Assume that a one-way function $f$ mapping $\Sigma_1^*$ to $\Sigma_2^*$ exists in $\flin$. Let $f^{-1}$ denote a multi-valued partial function defined as follows. For each string of the form $\track{y}{1^n}$, 
if $|y|\geq n$, then we define $f^{-1}\left(\track{y}{1^n}\right)
=\{x\#^{|y|-n}\mid |x|=n, f(x) =y\}$; otherwise, let $f^{-1}\left(\track{y}{1^n}\right)
=\{x\mid |x|=n, f(x) =y\}$. 
Note that $f^{-1}$ is length-preserving and belongs to $\nlinmv$. Lemma \ref{refinement} ensures the existence of a $\flin\mathrm{(partial)}$ function $g$ that is a refinement of $f^{-1}$. Consider the following 1TM $M$: on input $\track{y}{1^n}$, 
check if $\track{y}{1^n}\in \dom(g)$. If not, $M$ outputs any fixed string of length $n$ (\eg $0^{n}$). Otherwise, $M$ computes $g\left(\track{y}{1^n}\right)$ and outputs a string obtained from it by deleting the symbol $\#$. Since $\dom(g)$ is in $\dlin$, $M$ can be deterministic. Clearly, $M$ inverts $f$; that is, $f\left(M\left(\track{f(x)}{1^{|x|}}\right)\right) = f(x)$ for all inputs $x$. This contradicts the one-wayness of $f$. Therefore, $f$ cannot be one-way.  
\end{proof}

The third application concerns the advised class $\mathrm{REG}/n$. Similar to this class, we define $\dlin/\mathrm{lin}$ as the collection of all languages $A$ such that there are a linear-time deterministic 1TM $M$, an advice function $h$, and a constant $c\geq1$ for which (i) $|h(n)|\leq cn+c$ for any number $n\in\nat$, and (ii) for every $x$, $x\in A$ iff $\track{x}{h(|x|)}\in L(M)$.  Now, we can prove that the two classes 
$\mathrm{REG}/n$ and $\dlin/\mathrm{lin}$ coincide.

\begin{proposition}
$\mathrm{REG}/n = \dlin/\mathrm{lin}$.
\end{proposition}

\begin{proof}
The inclusion $\mathrm{REG}/n\subseteq \dlin/\mathrm{lin}$ is obvious. Now, we want to show that $\dlin/lin\subseteq \mathrm{REG}/n$. Let $A$ be any language, over alphabet $\Sigma$, in $\dlin/\mathrm{lin}$. Without loss of generality, we can take a linear-time deterministic 1TM $M$ and an advice function $h$ satisfying that (i) $n\leq |h(n)|\leq cn$ for any number $n\in\nat$ and (ii) for every $x$, $x\in A$ iff $\track{x}{h(|x|)}\in L(M)$. 
For simplicity, we assume that an alphabet for our advice strings is different from $\Sigma$. 

Let $x$ be any input of length $n$. Initially, the tape of $M$ consists of the string $\track{x\#^{|h(n)|-n}}{h(n)}$. A folding machine $N$, induced from $M$, starts with its own input, say $cont(x,h(n))$, which is obtained 
by folding the tape content $\track{x\#^{|h(n)|-n}}{h(n)}$. 
{}From this input string $cont(x,h(n))$, we can construct another string simply by deleting all symbols in $\Sigma$. Since this new string does not include $x$, we denote it by $h'(n)$. Note that $|h'(n)|=n$.  

Let us describe a new deterministic 1TM $M'$ that behaves as follows. On input $\track{x}{h'(|x|)}$, $M'$ first modifies the input to $cont(x,h(|x|))$ in linear time and then simulates the folding machine $N$ using this new string as an input. Obviously, for every string $x$, $x\in A$ iff $M'$ accepts $\track{x}{h'(|x|)}$. Since $M'$ runs in linear time using only its input area, we can translate $M'$ into its equivalent deterministic finite automaton. Therefore, we can conclude that $A$ belongs to $\mathrm{REG}/n$.
\end{proof}

\section{Alternating Computation}\label{sec:alternating}

Chandra, Kozen, and Stockmeyer \cite{CKS81} introduced
the concept of alternating Turing machines
as a natural extension of nondeterministic Turing machines.
We first give a general description of an alternating 1TM 
using our strong definition of running time.
An {\em alternating 1TM} is
defined similar to a nondeterministic 1TM except that
its internal states are all labeled with symbols in $\{\exists,\forall\}$,
where $\exists$ reads ``existential'' and $\forall$ reads ``universal''
(this labeling is done by a fixed function that
maps the set of internal states to $\{\exists,\forall\}$).
All the nodes of a computation tree are evaluated inductively
as either $T$ (true) or $F$ (false) from the leaves to the root
according to the label of an internal state given in each node 
in the following recursive fashion. 
A leaf is evaluated $T$ if and only if it is in the accepting state.
An internal node labeled with symbol $\exists$ is evaluated $T$
if and only if at least one of its children is evaluated $T$.
An internal node labeled with symbol $\forall$ is evaluated $T$
if and only if all of its children are evaluated $T$.
An alternating 1TM $M$ {\em accepts} input $x$ exactly
when the root of the computation tree of $M$ on $x$ is evaluated $T$.

The {\em $k$-alternation} means that 
the number of the times when internal states change 
between different labels is at most $k-1$
along every computation path.
For instance,
a nondeterministic Turing machine can be viewed as
an alternating Turing machine whose internal states are all 
labeled $\exists$, and therefore it has $1$-alternation.
Let $k$ and $T$ be any functions from $\nat$ to $\nat$ satisfying that
$k(n)\leq T(n)$ for all $n\in\nat$.
The notation $1\mbox{-}\sigmatime{k(n)}{T(n)}$
($1\mbox{-}\pitime{k(n)}{T(n)}$, resp.)
expresses the collection of all languages recognized by certain
$T(n)$-time alternating 1TMs with {\em at most} $k(n)$-alternation starting
with an $\exists$-state (a $\forall$-state, resp.).
For any given language $A\in1\mbox{-}\sigmatime{k(n)}{T(n)}$,
take a $T(n)$-time alternating 1TM $M$ that
recognizes $A$ 
with at most $k(n)$-alternation starting with an $\exists$-state.
Define $\overline{M}$ to be the one obtained from $M$ by exchanging
$\forall$-states and $\exists$-states and
swapping an accepting state and a rejecting state.
It follows that
$\overline{M}$ is a $T(n)$-time alternating 1TM
with at most $k(n)$-alternation starting with a $\forall$-state. Clearly, $\overline{M}$ recognizes $\overline{A}$.
Thus,
$\co1\mbox{-}\sigmatime{k(n)}{T(n)}\subseteq 1\mbox{-}\pitime{k(n)}{T(n)}$.
Similarly, we have
$\co1\mbox{-}\pitime{k(n)}{T(n)}\subseteq 1\mbox{-}\sigmatime{k(n)}{T(n)}$,
and hence $1\mbox{-}\pitime{k(n)}{T(n)}=\co1\mbox{-}\sigmatime{k(n)}{T(n)}$.
Given a set ${\cal T}$ of time-bounding functions,
$1\mbox{-}\sigmatime{k(n)}{{\cal T}}$
($1\mbox{-}\pitime{k(n)}{{\cal T}}$, resp.) stands for the union of all sets
$1\mbox{-}\sigmatime{k(n)}{T(n)}$ ($1\mbox{-}\pitime{k(n)}{T(n)}$, resp.)~%
over all functions $T$ in ${\cal T}$.
In particular,
we write $\sigmalinh{k(n)}$ ($\pilinh{k(n)}$, resp.)
for $1\mbox{-}\sigmatime{k(n)}{O(n)}$ ($1\mbox{-}\pitime{k(n)}{O(n)}$, resp.).

Of our particular interest are alternating 1TMs
with a constant number of alternations.
When $k$ is a constant in $\nat^{+}$,
it clearly holds that $\pilinh{k}=\co\sigmalinh{k}$ and
$\reg \subseteq \sigmalinh{k}\cap\pilinh{k}\subseteq
\sigmalinh{k}\cup\pilinh{k}\subseteq \alin$.
Similarly, we define the complexity class 
 $\deltalinh{k}$ by 
$\deltalinh{k+1}=\dlin_m^{\sigmalinh{k}}$ for every index $k\in\nat$.
Now, we generalize the earlier collapse result $\nlin = \reg$ 
and prove that three complexity classes
$\sigmalinh{k}$, $\pilinh{k}$, and $\deltalinh{k}$
all collapse to $\reg$.

\begin{theorem}\label{theorem:alternating}
$\reg=\bigcup_{k\in\nat^+}\sigmalinh{k}
=\bigcup_{k\in\nat^+}\pilinh{k}
=\bigcup_{k\in\nat^+}\deltalinh{k}$.
\end{theorem}

Theorem \ref{theorem:alternating} follows from Proposition
\ref{lemma:REG-closure} and the following lemma.
In this lemma, we show that 
alternation can be viewed as
an application of many-one 
$1\mbox{-}\mathrm{NLIN}$-reductions.

\begin{lemma}\label{lemma:NLIN-reduction}
For every number $k\in\nat^+$, $\sigmalinh{k+1}=\nlin_{m}^{\pilinh{k}}$.
\end{lemma}

The proof of Theorem \ref{theorem:alternating} proceeds by induction on $k$.
The base case $k=1$, \ie $\pilinh{1}=\sigmalinh{1}=\reg$,
is already shown in Theorem \ref{ntime}
since an alternating 1TM with 1-alternation starting with an
$\exists$-state is identical to a nondeterministic 1TM.
The induction step $k>1$ is carried out as follows.
Assume that $A$ is in $\sigmalinh{k}$.
Lemma \ref{lemma:NLIN-reduction} yields the existence of
a set $B\in\pilinh{k-1}$ satisfying that $A\in\nlin_{m}^{B}$.
By the induction hypothesis,
$B$ falls into $\reg$ and hence $A$ is in $\nlin_{m}^{\reg}$, 
which is obviously $\reg$.
Since $\pilinh{k}=\co\sigmalinh{k}$, $\pilinh{k}$ also collapses to $\reg$.
Similarly, from the inclusion $\deltalinh{k} \subseteq \sigmalinh{k}$, it follows that  $\deltalinh{k}=\reg$.

\begin{proofof}{Lemma \ref{lemma:NLIN-reduction}}
($\subseteq$-direction)
Let $A$ be any language  
in $\sigmalinh{k+1}$, where $k\geq1$, over alphabet $\Sigma_1$. 
Take a linear-time alternating 1TM $M$ with at most $k$-alternation that
recognizes $A$. Without loss of generality,
we can assume that $M$ never visits the cell indexed $-1$
since, otherwise, we can ``fold'' a computation into two tracks,
in which the first track simulates the tape region of nonnegative indices,
and the second track simulates the tape region of negative indices.

First, we define a linear-time nondeterministic 1TM $M'$ that simulates $M$ during its first alternation.
Let $\#$ be a new symbol not in $\Sigma_1$. 
On input $x$, $M'$ marks the start cell and then starts simulating $M$. 
During this simulation, whenever $M$ writes a blank symbol,
$M'$ replaces it with $\#$.
When $M$ enters the first $\forall$-state $p$, $M'$ first marks 
the currently scanning cell
(by changing its tape symbol $a$ to the new symbol $\track{a}{p}$),
moves its tape head back to the start cell, 
and finally erases the mark at the start cell.
It is important to note that
$M'$ has only one block of 
at least $|x|$ non-blank symbols on its tape when it halts.
Let $\Sigma$ consist of all symbols of the form $\track{a}{p}$, where 
$a$ is any symbol in $\Sigma_1$ and $p$ is any $\forall$-state.

Next, we define another alternating 1TM $N$ as follows.
Let $\Sigma_2=\Sigma_1\cup\Sigma\cup\{\#\}$ be a new alphabet for $N$.
On input $y$ in $\Sigma_2^*$, $N$ changes all $\#$s to the blank symbol,
finds on the tape the leftmost cell that contains a symbol, 
say $\track{a}{p}$, from $\Sigma$, and changes it back to $a$.
At the same time, $M'$ recovers the $\forall$-state $p$ as well.
By starting with this internal state $p$ of $M$, $N$ simulates $M$ 
step by step.
Finally, the desired set $B$ is defined 
to include all input strings accepted by $N$. 
Obviously, $B$ is in $\pilinh{k-1}$.
By their definitions, $M'$ many-one {1-NLIN}-reduces $A$ to  
$B$.

($\supseteq$-direction)
Assume that $A$ is in $\nlin_{m}^{\pilinh{k}}$; namely, 
$A$ is many-one {1-NLIN}-reducible to $B$ via a reduction machine $N$,
where $B$ is a certain set in $\pilinh{k}$.
Choose a linear-time alternating 1TM $M$ that
recognizes $B$ with $k$-alternation starting with a $\forall$-state.
Now, let us define $N'$ as follows:
on input $x$, simulate $N$, and, when $N$ eventually halts, simulate $M$ on the same input/work tape.
Clearly, $N'$ runs in linear time since so do $M$ and $N$.
It is also easy to show that $N'$ recognizes $A$ with $(k+1)$-alternation
starting with an $\exists$-state.
Thus, $A$ belongs to $\sigmalinh{k+1}$.
\end{proofof}

The collapse of the hierarchy of alternating complexity classes with constant-alternation depends on our strong definition of nondeterministic running time. By contrast, when the linear-time alternating class is defined with a {\em weak definition} of running time (\eg the length of the shortest accepting path if one exists, and $1$ otherwise),  the language $L=\{x\#y\mid x,y\in\Sigma^*,\text{ $y$ is the binary representation
of $|x|$}\}$ can separate this alternating class from $\reg$. (See \cite{PPR80} also \cite{BDG90}.) 

\section{Probabilistic Computation}
\label{sec:probabilistic}

Probabilistic (or randomized) computation 
has been proven to be essential to
many applications in computer science. 
Since as early as the 1950s, probabilistic extensions of
deterministic Turing machines have been 
studied from theoretical interest as well as for practical applications. 
This paper adopts Gill's model of 
probabilistic Turing machines with flipping 
{\em fair coins} \cite{Gil77}. Formally, we define a 
{\em probabilistic 1TM} as a 
nondeterministic 1TM that has {\em at most} two
nondeterministic choices at each step, 
which is referred to as a {\em coin
toss} (or {\em coin flip}) whenever there are exactly two choices. 
Each fair coin toss is made with probability exactly 
$1/2$. Instead of taking an expected running 
time, we define a
probabilistic 1TM $M$ to be 
{\em $T(n)$-time bounded} if, for each string 
$x$, all computation paths of $M$ on the input $x$ 
have length at most
$T(|x|)$. This definition reflects our strong definition of running time.
 The probability associated with each computation 
path $s$ equals
$2^{-m}$, where $m$ is the number of coin tosses 
along the path $s$. The
{\em acceptance probability} of $M$ on the 
input $x$, denoted $p_{M}(x)$,
is the sum of the probabilities of all accepting 
computation paths. For any language $L$, we
say that $M$ {\em recognizes $L$ with error 
probability at most $\varepsilon$}
if, for every $x$,
(i) if $x\in L$, then $p_{M}(x) \ge 1-\varepsilon$;
and (ii) if $x\not\in L$, then $p_{M}(x)\le\varepsilon$.

We begin with a key lemma, which is a probabilistic 
version of Lemma
\ref{nbound-regular}. Ka\c{n}eps and Freivalds 
\cite{KF90}, following
Rabin's \cite{Rab63} result, proved a similar 
result for probabilistic
finite automata.

\begin{lemma}\label{lemma:prop-error}
Let $L$ be any language and let $M$ be any
probabilistic 1TM that recognizes $L$ with error 
probability at most
$\varepsilon(n)$, where $0\leq \varepsilon(n)<1/2$ 
for all numbers $n\in\nat$. For each number $n\in \nat$, 
let $S_n$ be the union, over all strings $x$ of length 
at most $n$, of the sets of all
crossing sequences at any critical-boundary 
of $x$ along any accepting
computation path of $M$ on $x$.  Then,
$N_{L}(n)\leq 2^{|S_n|\ceilings{|S_n|/\delta(n)}}$ 
for all
$n\in\nat$, where $\delta(n)=1/2-\varepsilon(n)$.
\end{lemma}

\begin{proof}
Fix $n\in\nat$ arbitrarily.  For every string
$x\in\Sigma^{\leq n}$ and every crossing 
sequence $v\in S_n$, let $w_{l}(x| v)$  
be the sum, over all $z$ with $|xz|\leq n$, of all
probabilities of the coin tosses made during the tape head
staying in the left-side region of the 
right-boundary of $x$ along any
accepting computation path of $M$ on the input $xz$. 
Similarly, for every $z\in\Sigma^{\leq n}$ and every 
$v\in S_n$, let $w_{r}(v| z)$  
be the sum, over all $x$ with $|xz|\leq n$, of all
probabilities of the coin tosses made during the tape head
staying in the right-side region of the left-boundary 
of $z$ along any
accepting computation path of $M$ on the input $xz$. 
By these two definitions, it follows that  
$0\leq w_{l}(x|v),w_{r}(v|z)\leq 1$.  
The key observation is that the
acceptance probability of $M$ on the input 
$xz$ with $|xz|\leq n$ equals
$\sum_{v\in S_n}w_{l}(x|v)w_{r}(v|z)$.
 
Now, we say that $x$ {\em $n$-supports} 
$(i,v)$ if $|x|\leq n$,
$i\in [0,\ceilings{|S_n|/\delta(n)}-1]_{\integer}$, 
$v\in S_n$, and
$i\cdot \delta(n)/|S_n| \leq w_{l}(x|v) \leq
(i+1)\delta(n)/|S_n|$.
Define the support set $\Supp{n}{x}=\{(i,v)\mid 
\mbox{ $x$ $n$-supports $(i,v)$}\}$. 
We first show that, for every $x,y,z$ with 
$|xz|\leq n$ and $|yz|\leq n$, if $xz\in
L$ and $\Supp{n}{x}=\Supp{n}{y}$, then $yz\in L$. 
This is shown as follows. Since
$\Supp{n}{x}=\Supp{n}{y}$, $|w_{l}(x|v)-w_{l}(y|v)|\leq
\delta(n)/|S_n|$ for all crossing sequences $v\in S_n$. Thus,
$|p_{M}(xz) -p_{M}(yz)| = 
\left|\sum_{v\in S_n}(w_{l}(x| v)- 
w_{l}(y| v))\cdot w_r(v| z)\right| 
\leq \sum_{v\in S_n} \left|w_{l}(x| v)- w_{l}(y| v)\right| 
\leq \sum_{v\in S_n} \delta(n)/|S_n| = \delta(n)$.
Since $xz\in L$, we obtain 
$p_{M}(xz)\geq 1 - \varepsilon(n)$, which yields
$p_{M}(yz) > \varepsilon(n)$. Hence, we obtain $yz\in L$.

Note that $N_{L}(n)$ is bounded above by 
the number of distinct
$\Supp{n}{x}$'s for all strings $x\in\Sigma^{\leq n}$. 
Therefore, $N_{L}(n)$ is at most
$2^{|S_n|\ceilings{|S_n|/\delta(n)}}$, as requested.
\end{proof}

Let us focus our attention on the case 
where the error probability of a probabilistic 
1TM is bounded away from $1/2$.  
For each language
$L$ and any probabilistic 1TM $M$, we say that 
$M$ {\em recognizes $L$ 
with bounded-error
probability} if there exists a constant 
$\varepsilon>0$ such that $M$
recognizes $L$ with error probability at 
most $1/2 -\varepsilon$.
We define $1\mbox{-}\bptime{T(n)}$ as the 
collection of all languages 
recognized by $T(n)$-time probabilistic 1TM 
with bounded error
probability. We also define 
$1\mbox{-}\bptime{{\cal T}}$ for any set $\TT$ 
of time-bounding functions.
The {\em one-tape bounded-error probabilistic
linear-time} class $\bplin$ is $1\mbox{-}\bptime{O(n)}$.

Consider any language $L$ recognized by a probabilistic 
1TM $M$ with bounded-error
probability in time $o(n\log n)$. Lemma \ref{nlogn-bound} implies that 
the number of all crossing sequences of $M$ is upper-bounded
by a certain constant independent of its input. It thus follows from Lemma \ref{lemma:prop-error} that $N_L(n)$ is bounded above by an exponential function of the machine's error bound $\varepsilon(n)$. Since $\varepsilon(n)$ is bounded away from 1/2, 
we obtain $N_{L}(n)=O(1)$, which yields the regularity of $L$. Therefore, 
 $\reg=1\mbox{-}\bptime{o(n\log n)}$. The separation  $\reg \neq 1\mbox{-}\bptime{O(n\log n)}$ follows from Proposition \ref{prop:Hennie}.

\begin{theorem}\label{bptime}
  $\reg=1\mbox{-}\bptime{o(n\log n)} 
\subsetneqq 1\mbox{-}\bptime{O(n\log n)}$.
\end{theorem}

Leaving from bounded-error probabilistic computation,
we hereafter concentrate on unbounded-error probabilistic computation.
We define $\plin$ to be the collection of all
languages of the form 
$\{x\in\Sigma^* \mid p_M(x)>1/2\}$ for certain 
linear-time probabilistic 1TMs $M$. Different from
$\bplin$, $\plin$ does not collapse to $\reg$ because
the non-regular set $L_{>}=\{a^mb^n\mid m>n\}$ is in
$\plin$. 

The following theorem establishes a 1TM-characterization of $\slrat$. A similar characterization of $\slrat$ was given by Ka\c{n}eps \cite{Kan89} in terms of one-head two-way probabilistic automata with rational transition probabilities. For simplicity, 
we write $\synplin$ for the subset of $\plin$ 
defined by linear-time probabilistic 1TMs that 
are particularly {\em synchronous}. 

\begin{theorem}\label{1plin-rstcl-complete}
$\plin = \synplin = \slrat$.
\end{theorem}

The class $\plin$ is easily shown to be
closed under complementation
and symmetric difference, where the {\em symmetric difference} 
between two sets $A$ and $B$ is $(A-B)\cup(B-A)$. These properties also result from Theorem \ref{1plin-rstcl-complete} using the corresponding properties of $\slrat$.

Now, let us prove Theorem \ref{1plin-rstcl-complete}. The theorem  
follows from two key lemmas: Lemmas \ref{rstcl-subset-1synplin}
and \ref{1plin-subset-rgstcl}. We begin with 
Lemma \ref{rstcl-subset-1synplin} whose proof is based on a simple simulation
of rational 1PFAs by synchronous probabilistic 1TMs.

\begin{lemma}\label{rstcl-subset-1synplin}
$\slrat \subseteq \synplin$.
\end{lemma}

\begin{proof}
Let $L$ be any language in $\slrat$. There exists a rational 1PFA $N =(S, \Sigma,\pi,\{T(\sigma)\,|\,\sigma\in\Sigma\}, \eta)$ with a rational cut point for $L$. Without loss of generality, we can assume that (i) $L=L(N,1/2)$, (ii) $S=[1,s]_{\integer}$ for a certain number $s\in\nat^{+}$, (iii) one entry of $\pi$ equals $1$, and
(iv) there is a positive integer $d$ satisfying the 
following property: for any symbol $\sigma\in\Sigma$ and any pair $i,j\in S$,
the $(i,j)$-entry of the matrix $T(\sigma)$, denoted $T(\sigma)_{i,j}$, 
is of the form $r_{i,j}(\sigma)/d$ for a certain number $r_{i,j}(\sigma)\in\nat$. Write $F$ for the set of all the final states of $N$.

Our goal is to construct a synchronous probabilistic 1TM $M$ that simulates $N$ in linear time with unbounded error. The desired machine $M$ works as follows. In case where the input is the empty string $\lambda$, $M$ immediately enters $q_{acc}$ or $q_{rej}$ depending on $\lambda\in L$ or $\lambda\not\in L$, respectively. Henceforth, assuming that our input is not $\lambda$, we give an algorithmic description of $N$'s behavior. Choose an integer $m$ such that $2^{m-1}<d\leq 2^{m}$. First, we repeat phases 1)-2) until $M$ finishes scanning all input symbols. Initially, $M$ sets its {\em decision value} to be $-1$.

\ms

1) In scanning a symbol $\sigma$, $M$ first generates 
$2^{m}$ branches by tossing exactly $m$ fair coins 
without moving its head. The (lexicographically) first $d$ branches 
are called {\em useful}; the other branches are called {\em useless}.
The useful branches are used for 
the simulation of a single step of $N$ in phase 2. 

2) First, we consider the case where the current decision value is $-1$. In the following manner, $M$ simulates a single step of $N$'s moves. Assume that $N$ is in internal state $i$. Note that, for any choice $j\in S$, $N$ changes the internal state $i$ to $j$ with the transition probability $T(\sigma)_{i,j}$ $(=r_{i,j}(\sigma)/d)$ while scanning the symbol $\sigma$. To simulate such a transition, we choose exactly $r_{i,j}(\sigma)$ branches out of the useful $d$ branches and then follow the same transition of $N$. More precisely, along the $\ell$th branch generated by the coin tosses made in phase 1, $M$ simulates $N$'s transition from the internal state $i$ to  $j$ if $\sum_{k=1}^{j-1}r_{i,k}(\sigma)< \ell \leq \sum_{k=1}^{j}r_{i,k}(\sigma)$. We then force $M$'s head to move to the right-adjacent cell. Along the useless branches, $M$ tosses a fair coin, remembers its outcome (either $0$ or $1$) as a new decision value, and moves its head rightward. If the current decision value is not $-1$, then we simply force $M$'s head to step to the right.  

3) When $M$ finishes reading the 
entire input, its head must sit in the first blank cell. 
With the decision value $-1$, if $N$ reaches a final state in $F$, then $M$ enters $q_{acc}$; otherwise, $M$ enters $q_{rej}$. If the decision value is either $0$ or $1$, $M$ enters $q_{rej}$ or $q_{acc}$, respectively. 
This completes the description of $M$.

\ms

By our simulation, the acceptance probability of $N$ on input $x$ is greater  than $1/2$ iff the acceptance probability of $M$ on the same input is more than $1/2$.
Moreover, our simulation makes $M$'s computation 
paths terminate all at once.
Therefore, $L$ is in $\synplin$ via $M$. 
\end{proof}

The following lemma complements Lemma 
\ref{rstcl-subset-1synplin}.  

\begin{lemma}\label{1plin-subset-rgstcl}
For any probabilistic 1TM $M$ running in linear time,
 there exists a rational 1GPFA $N$ 
such that $p_{N}(x)=p_{M}(x)$ for any input $x$.
\end{lemma}

To lead to the desired consequence $\plin\subseteq \slrat$, take a language $L$ in $\plin$ and consider any linear-time probabilistic 1TM $M$ that recognizes $L$ with unbounded-error probability. Lemma \ref{1plin-subset-rgstcl} guarantees the existence of a rational 1GPFA $N$ for which $L=L(N,1/2)$. Hence, $L$ is in $\gslrat$, which is known to equal $\slrat$. With Lemmas \ref{rstcl-subset-1synplin}, we therefore obtain Theorem \ref{1plin-rstcl-complete}.

\begin{proofof}{Lemma \ref{1plin-subset-rgstcl}}
Let $M=(Q,\Sigma,\Gamma,\delta,q_0,q_{acc},q_{rej})$ be
any linear-time probabilistic 1TM. In this proof, we need the folding machine $M'$ constructed from $M$.
To simplify the proof, we further modify $M'$ as follows. When $M'$ halts in a certain halting state, we force its head to move rightward and cross the left-boundary of the original input by entering the same halting state as $M$ does. Note that the accepting probability of this modified machine is the same as that of $M$. For notational simplicity, we use the notation $M$ to denote this modified machine. For this new machine $M$, the crossing sequence at the left-boundary of any input should be $v_0=()$ and the crossing sequence at the right-boundary of any input is $v_f=(q_1,q_2,q_{acc})$ along every accepting computation path of $M$. 

We wish to construct a rational 1GPFA $N = (S,
\Sigma,\pi,\{T(\sigma)\,|\,\sigma\in\Sigma\},\eta)$ 
satisfying that $p_N(x) = p_M(x)$ for all inputs $x\in\Sigma^*$. 
The desired automaton $N$ is defined in the following manner.  
Let $S$ denote the set of all crossing sequences of 
$M$. It follows from Lemma \ref{nlogn-bound} 
that $S$ is a finite set.
Let $\sigma$ be an arbitrary symbol in $\Sigma$. 
For any pair $(u,v)$ of elements in $S$, 
we define 
$P(u;\sigma;v)$ to be the {\em probability} of the 
following event ${\cal E}$. 

\begin{quote}
{\sf Event ${\cal E}$: 
Consider any computation tree of $M$ where $M$ 
starts on input $y\sigma z$ with the tape head initially 
scanning the left-most symbol of the input $y\sigma z$
for a certain pair $(y,z)$ of strings.  
In a certain computation path of this computation tree, 
(i) $u$ coincides with the crossing sequence at the 
left-boundary of $\sigma$, (ii) $v$ is the crossing 
sequence at the right-boundary of $\sigma$, and (iii) 
$u\rightarrow_{\sigma}v$.
}
\end{quote}

Clearly, $P(u;\sigma;v)$ is a dyadic rational number 
since $M$ flips only fair coins. Let 
$x=\sigma_1\dotsm\sigma_n$ be any nonempty input string, where 
each $\sigma_i$ is in $\Sigma$. By the correspondence between a series of crossing sequences and a computation path, the acceptance probability $p_{M}(x)$ equals $\sum_{\vec{v}} \prod_{i=1}^{n}
  P(v_{i-1};\sigma_i;v_i)$, where the sum is taken over all sequences 
$\vec{v}=(v_0,v_1,\dots,v_{n})$ from $S$ 
with $v_n= v_f$. 
For each tape symbol 
$\sigma\in\Sigma$, define $T(\sigma)$ to be the $|S|\times |S|$
matrix whose $(u,v)$-element is 
$P(u;\sigma;v)$ for any pair $u,v\in S$. 
The row vector $\pi$ has $1$ or $0$ in the 
$v$th column if $v = v_0$ or $v\neq v_0$, respectively, for any $v\in S$.
Letting $F=\{v_0,v_f\}$ if $\lambda\in L$ and 
$F=\{v_f\}$ otherwise,
we define $\eta$ to be the column vector whose
$v$th component is 1 or 0 if $v\in F$ or $v\not\in F$, respectively. Thus, we have $p_N(x) =  \pi\,T(x)\,\eta$ for every input string $x$. 

By the above definition of $N$, it is not difficult to 
verify that,  for each input $x$, 
$p_{N}(x) = \pi\,T(x)\,\eta = \sum_{\vec{v}} \prod_{i=1}^{n}
  P(v_{i-1};\sigma_i;v_i) = p_{M}(x)$, as requested.
\end{proofof}

Macarie \cite{Mac98} showed the proper containment
$\slrat\subsetneqq\dl$, where 
$\dl$ is the class of all languages recognized by
multiple-tape deterministic Turing machines, 
with a read-only input tape and multiple 
read/write work-tapes, which uses $O(\log n)$ 
tape-space on all the tapes
except for the input tape (and halting eventually on all inputs).
We thus obtain the following consequence of Theorem \ref{1plin-rstcl-complete}.
Note that $\dl\nsubseteq\cfl$ since, for instance, 
$L_{3eq}=\{a^nb^nc^n\mid n\in\nat\} \in \dl - \cfl$. 

\begin{corollary}\label{corollary:DL}
$\plin\subsetneqq\dl$,
$\reg/n\nsubseteq\dl$ and $\dl\nsubseteq\reg/n$.
\end{corollary}

\begin{proof}
The proper inclusion $\plin\subsetneqq\dl$ follows from 
Theorem \ref{1plin-rstcl-complete} as well as the fact that $\slrat\subsetneqq\dl$.
Since $\reg/n$ contains all non-recursive tally languages, 
it immediately follows that  $\reg/n \nsubseteq\dl$.
To prove that $\dl\nsubseteq\reg/n$,
we use the language $Equal = \{ w\in\{0,1\}^* \mid \#_0(w)=\#_1(w) \}$.
While Lemma \ref{non-reg-advice} places
$Equal$ outside of $\reg/n$,
$Equal$ obviously belongs to $\dl$. 
We therefore obtain the last separation $\dl\nsubseteq\reg/n$. 
\end{proof}

Theorem \ref{1plin-rstcl-complete} also provides us with 
separations among three complexity classes $\reg/n$, $\cfl$, and $\plin$. 
We see these separation results in the next proposition. 

\begin{proposition}\label{proposition:plin}
$\cfl\cap \reg/n \nsubseteq \plin$, 
$\cfl \nsubseteq \plin\cup\reg/n$, and
$\reg/n \nsubseteq \cfl\cup\plin$.
\end{proposition}

Earlier, Nasu and Honda \cite{NH71} found
a context-free language not in $\slrat$. More precisely,
they introduced the context-free language
$L_{NH}=\{a^iba^{j_1}b\dots ba^{j_r}b\mid
r\in\nat^+\ \&\ i,j_1,\dots,j_r\in\nat\ \&\ i= \sum_{k=1}^{\ell}j_k 
\text{ for a certain number $\ell\in[1,r]_{\integer}$}\}$
and showed that, using the Cayley-Hamilton theorem, 
$L_{NH}$ cannot belong to $\slrat$.
A similar technique can show that $\slrat$ does not contain 
the context-free language $Center= \{x1y\mid x,y\in\{0,1\}^*,|x|=|y|\}$.

\begin{proofof}{Proposition \ref{proposition:plin}}
It is easy to show that the context-free language $Center$ 
falls into $\reg/n$ by choosing advice of the form $0^n10^n$ 
whenever the length $|x1y|$ is odd. Since $Center\not\in \slrat$, 
the first separation follows from Theorem \ref{1plin-rstcl-complete}.

For the second separation, let us 
consider the context-free non-stochastic language $L_{NH}$.
It is enough to prove that $L_{NH}\notin\reg/n$. This can be done in a similar fashion as in the proof of Lemma \ref{non-reg-advice}. 
Let us fix our alphabet $\Sigma=\{a,b\}$. 
Assuming that $L_{NH}\in\reg/n$, choose a deterministic 
finite automaton $M=(Q,\Sigma,q_0,F)$ 
and an advice function $h$ mapping from $\nat$
to $\Sigma^*$ such that, for every
$x\in\Sigma^*$, $x\in L_{NH}$ if and only if $\track{x}{h(|x|)}\in B$.
Let $n = |Q|+2$. 
For each number $k\in [1,n-1]_{\integer}$, define
$y_k$ and $z_k$ to be the strings
$a^{n+k}ba^{n-k}b$ and $a^{2k}ba^{2(n-k)}b$ of length $2n+2$, respectively.
Obviously, $y_kz_k$ belongs to $L_{NH}$ since 
$|a^{n+k}| = |a^{n-k}|+|a^{2k}|$.
Choose two {\em distinct} indices
$k,l\in [1,n-1]_{\integer}$ such that
$M$ enters the same internal state after reading $y_k$ as well as $y_l$. 
Such a pair $(k,l)$ exists since $n-1>|Q|$.
It thus follows that $M$ should accept the input $\track{y_kz_l}{h(|y_kz_l|)}$.
This yields the membership $y_kz_l\in L_{NH}$.
Clearly, this contradicts the definition of $L_{NH}$.
Therefore, we conclude that $L_{NH}\not\in\reg/n$.

The third separation is rather obvious because $\reg/n$ contains a non-recursive language. 
\end{proofof}

\section{Counting Computation}\label{sec:counting}

Counting issues naturally arise in many fields of computer science.
For instance, the decision problem of determining whether 
there exists a Hamiltonian cycle in a given graph 
induces the problem of counting 
the number of such cycles. In the late 1970s, Valiant \cite{Val79} introduced the notion of {\em counting Turing machines} to study the complexity of counting. Our goal is to investigate the functions computed by one-tape linear-time counting Turing machines.

\subsection{Counting Functions}

A {\em counting 1TM} is a variant 
of a nondeterministic
1TM, which behaves like a nondeterministic 
1TM except that, when it
halts, we take the number of all accepting 
computation paths as the outcome of the machine. 
Let $\#M(x)$ denote the outcome of such a 
counting 1TM $M$ on input $x$. In this way, counting 1TMs can compute 
(partial) functions mapping strings to numbers. 
These functions are called {\em counting functions}.

Similar to Valiant's function class $\sharpp$ \cite{Val79}, 
we use the notation
$\sharplin$ (pronounced ``one sharp lin'') 
to denote the collection
of all total functions $f$, from $\Sigma^*$ to $\nat$, 
which are computed by certain linear-time counting
1TMs. This function class $\sharplin$ naturally includes 
$\flin$ by identifying any natural number $n$ 
with the $n$th string over alphabet $\Sigma$ 
(in the standard order) and by constructing
a linear-time nondeterministic 1TM, which branches off into 
$n$ computation paths, starting with the $n$th string as an input.
Another useful function class besides $\onesharplin$ is $\gaplin$, 
which is defined as the
class of all total functions whose values are the 
difference between the
number of accepting paths and the number of 
rejecting paths of
linear-time nondeterministic 1TMs. Such functions 
are conventionally called {\em gap functions}. 

We can prove the following closure property. For convenience, write $\nlinmv_{dis}$ for the collection of all partial multi-valued functions computed by certain nondeterministic 1TMs whose valid computation paths always output distinct values. 

\begin{lemma}\label{gaplin-closure}
For any functions $f,g$ in $\gaplin$ and any function $h$ in $\nlinmv_{dis}$, 
the following functions all belong to $\gaplin$: 
$f\cdot g$, $f+g$, $f-g$, and 
$\lambda x.\sum_{y\in h(x)}f(\track{x}{y})$.
\end{lemma}

\begin{proof}
We show the lemma only for the last function because the other cases are easily shown. For any two functions $f\in\gaplin$ and $h\in\nlinmv$, let $k(x)=\sum_{y\in h(x)}f(\track{x}{y})$ for each input string $x$. Take a nondeterministic 1TM $M_h$ computing $h$ in linear time and also a linear-time counting 1TM $M_f$ that computes $f$. Consider the counting 1TM $N$ that behaves as follows. On input $x$, $N$ produces $\track{x}{x}$ in the tape and runs $M_h$ using only the lower track. When $M_h$ halts, by our output convention, it leaves $\track{x}{y}$, where $y\in h(x)$, in the output tape. Next, $N$ simulates $M_f$ on the input $\track{x}{y}$. It follows that, for every $x$, $\#N(x)$ equals $\sum_{y\in h(x)}\#M_f(\track{x}{y})$, which is exactly $k(x)$. Hence, $k$ belongs to $\gaplin$.
\end{proof}

The above closure property implies 
that, for instance,  $\gaplin = \sharplin - \sharplin$, where 
the notation $\FF-\GG$ stands for the set 
$\{f-g\mid f\in\FF,g\in\GG\}$.

\begin{lemma}\label{gap-characterization}
$\gaplin = \sharplin - \sharplin$.
\end{lemma}

\begin{proof}
Let $h$ be any function in $\sharplin-\sharplin$. Take two functions 
$f,g\in\sharplin$ satisfying $h=f-g$. Since $f,g\in\gaplin$, the difference function $f-g$ is also in $\gaplin$ by the closure property of $\gaplin$. Hence, $h$ is in $\gaplin$.

Conversely, let $h$ be any function in $\gaplin$. There exists a linear-time counting 1TM $M$ that witnesses $h$; that is, $h(x)=\#M(x)-\#\overline{M}(x)$ for all $x$, where $\#\overline{M}(x)$ denotes the number of all rejecting computation paths of $M$ on input $x$. Define $f(x)=\#M(x)$ and $g(x)=\#\overline{M}(x)$ for every $x$. Clearly, $f$ is in $\sharplin$. It is also easy to show that $g$ is in $\sharplin$. Since $h=f-g$, $h$ belongs to $\sharplin-\sharplin$.
\end{proof}

By $\gafrat$, we denote
the set of all acceptance functions of rational 
1GPFAs. Lemma \ref{gap-characterization} implies that $\gaplin$ 
is a proper subset of $\gafrat$.

\begin{lemma}\label{gaplin-gafrat}
$\gaplin\subsetneqq\gafrat$.
\end{lemma}

\begin{proof}
The inequality $\gaplin \neq \gafrat$ is obvious since 
certain functions in $\gafrat$ can 
output non-integer values whereas $\gaplin$ contains only integer-valued functions. 

For inclusion, we first note that $\sharplin \subseteq \gafrat$,
by re-defining the value $P(u;\sigma;v)$ in 
the proof of Lemma \ref{1plin-subset-rgstcl} 
to be the number of accepting computation paths
instead of probabilities.
By Lemma \ref{gap-characterization}, we can 
write any given function in $\gaplin$ as a difference $f_1 - f_2$ of
two functions $f_1$ and $f_2$ in $\sharplin$. The desired inclusion 
now follows from the fact that   
$\gafrat$ is closed under difference,  
in fact under any linear combinations \cite{Mac93}.
\end{proof}

Theorem \ref{1plin-rstcl-complete}
and Lemma \ref{gaplin-gafrat} build a bridge between
counting computation and unbounded-error 
probabilistic computation. Here, we show that $\plin$ can be characterized  in terms of $\gaplin$. 

\begin{proposition}\label{1plin-1gaplin-complete}
$\plin = \{A\mid \exists 
f\in\gaplin\;\,[A=\{x\mid f(x)>0\}]\}$.
\end{proposition}

\begin{proof}
Assume that $A=\{x\mid f(x)>0\}$ for a certain function $f$ in $\gaplin$.  Lemma \ref{gaplin-gafrat} puts $f$ into $\gafrat$. This makes $A$ fall into  $\gslrat$ with the cut point $0$. Since $\gslrat=\slrat$, Lemma \ref{rstcl-subset-1synplin} ensures that $A$ is indeed in $\plin$.

Conversely, let $A$ be any language in $\plin$. By Theorem \ref{1plin-rstcl-complete}, $A$ is also in $\slrat$. Following the proof of Lemma \ref{rstcl-subset-1synplin}, we can recognize $A$ in linear time by a certain synchronous probabilistic 1TM $M$ which tosses the equal number of fair coins on all computation paths on each input. Let $N$ be the machine obtained from $M$ by exchanging the roles of $q_{acc}$ and $q_{rej}$. Now, we view $M$ and $N$ as counting 1TMs. Define $f$ and $g$ to be the functions computed by the counting machines $M$ and $N$, respectively. It follows from the definition that, for every string $x$, $x\in A$ if and only if $f(x)>g(x)$. Since $f(x)>g(x)$ is equivalent to $(f-g)(x)>0$, we obtain the characterization $A=\{x\mid (f-g)(x)>0\}$. Since $f-g$ is in $\gaplin$ by Lemma \ref{gap-characterization}, this completes the proof.
\end{proof}

In comparison, $\nlin$ can be characterized in terms of $\sharplin$ as $\nlin = \{A\mid \exists 
f\in\sharplin\;\,[A=\{x\mid f(x)>0\}]\}$. 

We already know the inclusion $\flin\subseteq\sharplin$.
Furthermore, Proposition \ref{1plin-1gaplin-complete} yields the separation between $\flin$ and $\sharplin$. 

\begin{corollary}
$\flin\subsetneqq \sharplin$.
\end{corollary}

\begin{proof}
It is enough to show that if $\flin=\sharplin$ then $\dlin=\plin$. 
Since $\reg\neq \plin$, it immediately follows that 
$\flin\neq\sharplin$. Now, assume that $\flin=\sharplin$.
Let $A$ be any set in 
$\plin$ and we wish to show that $A$ is also in $\dlin$. By Lemma \ref{gap-characterization} and 
Proposition \ref{1plin-1gaplin-complete}, 
there exist two functions $f,g\in \sharplin$ for which 
$A=\{x\mid f(x)>g(x)\}$. By our assumption, 
these functions fall into $\flin$. Using deterministic 
1TMs that compute $f$ and $g$ in linear time, we 
can produce $\track{f(x)}{g(x)}$ in binary in linear time from $x$. 
We can further determine whether $f(x)>g(x)$ by comparing 
$f(x)$ and $g(x)$ bitwise. This gives a 
deterministic linear-time 1TM for $A$, and thus $A$ belongs to $\dlin$. 
Therefore,  we obtain $\dlin=\plin$, as requested.
\end{proof}

\subsection{Counting Complexity Classes of Languages}

The function classes $\sharplin$ and $\gaplin$ naturally induce quite 
useful counting complexity classes. First, we define the counting class $\splin$ 
to be the collection
of all languages whose characteristic 
functions
belong to $\gaplin$. Furthermore, let 
$\paritylin$ (pronounced ``one parity lin'') 
consist of all languages
of the form $\{x\in\Sigma^*\mid f(x)\equiv 1 \  (\mathrm{mod}\ 2)\}$ 
for certain functions $f$ in $\sharplin$. Obviously, 
$\reg \subseteq \splin\subseteq  \paritylin$.
More generally,
for each integer $k\ge 2$ and
each nonempty proper subset $R$ of $[0,k-1]_{\integer}$,
we define the counting class $\modlin{k}{R}$ to include all languages
of the form $\{x\in\Sigma^*\mid \exists\, r\in R\ [f(x)\equiv r \  (\mathrm{mod}\ k)]\}$
for certain functions $f\in \sharplin$. 
It follows that $\reg \subseteq  \modlin{k}{R}$ and, in particular,   $\paritylin=\modlin{2}{\{1\}}=\co\modlin{2}{\{0\}}$.

Despite their complex definitions, these counting classes are no more powerful than $\reg$. 
Hereafter, we wish to prove the collapse of 
these classes down to $\reg$. 
Our proof uses a crossing sequence argument.

\begin{theorem}\label{paritylin-reg}
$\reg = \splin = \paritylin = \modlin{k}{R}$
for every integer $k\ge 2$ and
every nonempty proper subset $R$ of $[0,k-1]_{\integer}$.
\end{theorem}

\begin{proof}
It suffices to show that $\modlin{k}{R}\subseteq \reg$.
This can be shown by modifying the 
proof of Lemma \ref{lemma:prop-error}. 
Here, we define $w_{l}(x|v)$ and $w_{r}(v|z)$ to denote
the number of accepting computation paths instead of 
the sum of probabilities.
Recall that $\#M(u)$ denotes the outcome
(\ie the number of accepting paths)
of $M$ on input $u$.
For every pair $x,z$ with $|xz|\leq n$, it holds that 
$\#M(xz) = \sum_{v\in S_n}w_{l}(x|v)w_{r}(v|z)$.

Now, let $\Supp{n}{x}$ be the set
$\{(r,v)\in[0,k-1]_{\integer}\times S_n\mid w_{l}(x|v)\equiv r \  (\mathrm{mod}\   k)\}$.
We wish to show that, for every $x,y,z$ with $|xz|\leq n$ and $|yz|\leq n$, if $xz\in L$ and $\Supp{n}{x}=\Supp{n}{y}$ then $yz\in L$. 
This is shown as follows.
Note that,
for each $v\in S_n$, there exists a unique number 
$r\in[0,k-1]_{\integer}$  satisfying that $(r,v)\in\Supp{n}{x}$.
This implies that $\#M(xz) - \#M(yz) = 
\sum_{v\in S_n}(w_{l}(x| v)- 
w_{l}(y| v))\cdot w_r(v| z)$ =
$\sum_{r=0}^{k-1}\sum_{(r,v)\in \Supp{n}{x}} (w_{l}(x| v)- w_{l}(y| v))\cdot w_r(v|z)$.
For each $(r,v)\in \Supp{n}{x}$, we have 
$w_{l}(x| v)\equiv r \  (\mathrm{mod}\   k)$ and
$w_{l}(y| v)\equiv r \  (\mathrm{mod}\   k)$
since $\Supp{n}{y}=\Supp{n}{x}$.
It therefore follows that $w_{l}(x| v)- w_{l}(y| v)\equiv 0 \  (\mathrm{mod}\   k)$. {}From this, we 
conclude that $\#M(xz) - \#M(yz) \equiv 0 \  (\mathrm{mod}\   k)$.
Since $\#M(xz)\equiv r_0 \  (\mathrm{mod}\   k)$ 
for a certain number $r_0\in R$,
we have $\#M(yz)\equiv r_0 \  (\mathrm{mod}\   k)$ 
for the same $r_0$. This means that $yz\in L$.

Notice that $N_{L}(n)$ is bounded above by the number of distinct sets 
$\Supp{n}{x}$ for all strings $x\in\Sigma^{\leq n}$.
As a consequence, $N_{L}(n)$ is upper-bounded by $2^{k|S_n|}$,
which is bounded above by a certain constant.
Therefore, $L$ belongs to $\reg$.
\end{proof}

We wish to show an immediate consequence of Theorem \ref{paritylin-reg}
regarding low sets. Note that the notion of many-one $\nlin$ reducibility 
can be further expanded into other complexity classes 
(such a complexity class is called {\em many-one relativizable}). We can naturally define the many-one relativized counting class $\sharplin_{m}^A$ relative to set $A$ as the collection of all single-valued total functions $f$ such that there exists a linear-time nondeterministic 1TM $M$ satisfying the following: on every input $x$, $M$ produces an output $y_p$ along each computation path $p$ and $f(x)$ equals the number of all computation paths $p$ for which $y_p\in A$. Similarly, the many-one relativized class $\gaplin_{m}^A$ is defined using the difference $|\{p\mid y_p\in A\}| - |\{p\mid y_p\not\in A\}|$. A language $A$ is called {\em many-one low} 
for a relativizable complexity class
$\CC$ of languages or of functions if 
$\CC_{m}^{A}\subseteq \CC$. A complexity class $\DD$ 
is {\em many-one low} for $\CC$ if
every set in $\DD$ is many-one low for $\CC$. 
We use the notation  $\low_{m}\CC$ to denote the 
collection of all languages
that are many-one low for $\CC$. For instance, we obtain 
$\low_{m}\nlin = \reg$ since $\nlin_{m}^{\reg}=\nlin$.

\begin{corollary}\label{low-gaplin}
$\reg = \low_{m}\sharplin = \low_{m}\gaplin$.
\end{corollary}

\begin{proof}
To prove the corollary, we first note that 
$\reg\subseteq \low_{m}\sharplin$ since $\sharplin^{\reg}_{m}=\sharplin$. Conversely, 
for any set $A$ in $\low_{m}\gaplin$, since 
$\chi_{A}\in\sharplin_{m}^{A}\subseteq \gaplin_{m}^{A}$, 
it follows that $\chi_{A}\in \gaplin$. Thus, 
$A$ is in $\splin$, which equals $\reg$ by 
Theorem \ref{paritylin-reg}. Therefore, 
$\low_{m}\gaplin\subseteq \reg$. 
\end{proof}

We further introduce another counting class
$\cequallin$ (pronounced ``one C equal lin'') as 
the collection of all languages of the form 
$\{x\mid f(x)=0\}$ for certain functions $f$ 
in $\gaplin$. This class $\cequallin$ properly contains 
$\reg$ because
the non-regular language $L_{eq} = \{0^n1^n \mid n\in\nat \}$ clearly belongs to
$\cequallin$. Using the closure property of $\gaplin$, we can easily show that $\cequallin$ is closed under intersection and union. This is shown as follows. Let $A=\{x\mid f(x)=0\}$ and $B=\{x\mid g(x)=0\}$ for certain functions $f,g\in\gaplin$. Obviously, $A\cap B=\{x\mid f^2(x)+g^2(x)=0\}$ and $A\cup B=\{x\mid f(x)g(x)=0\}$. By Lemma \ref{gaplin-closure}, $A\cap B$ and $A\cup B$ are in $\cequallin$.

We wish to show robustness of the complexity class $\cequallin$. For comparison, we introduce $\syncequallin$ as the 
subset of $\cequallin$ defined by linear-time {\em synchronous} counting 1TMs. 
A similar argument to the proof of Lemma \ref{1plin-subset-rgstcl}
yields the simple containment $\cequallin \subseteq \slrateq$.
Moreover,
similar to Lemma \ref{rstcl-subset-1synplin}, 
we can prove that $\slrateq \subseteq \syncequallin$. 
Therefore, we obtain the following characterization of $\slrateq$.

\begin{theorem}\label{cequallin-slrateq}
$\cequallin = \syncequallin = \slrateq$
\end{theorem}

Earlier, Turakainen \cite{Tur69a} proved that
$\slrat$ is closed under complementation and that 
$\slrateq$ is properly included in $\slrat$. 
Symmetrically, $\co\slrateq$ is also properly included in $\slrat$.
In addition, Dieu \cite{Die71} showed 
that $\slrateq$ is not closed under complementation.
Dieu's argument can also work to show that the language
$L_{\ge}=\{a^mb^n\mid m\ge n \}$ cannot belong to  $\slrateq\cup\co\slrateq$.
Since $L_{\ge}$ belongs to $\plin$, 
Theorem \ref{cequallin-slrateq} immediately leads to 
the following separation results.

\begin{corollary}\label{cfl-pequallin}
$\cequallin \nsubseteq \co\cequallin$,
$\co\cequallin \nsubseteq \cequallin$,
and $\cequallin\cup\co\cequallin \subsetneqq \plin$.
\end{corollary}

In the next lemma, we briefly summarize basic relationships 
between $\cequallin$ and $\plin$. 

\begin{lemma}\label{cequallin-nlin-red}
$\cequallin \subseteq \plin \subseteq \nlin_{m}^{\cequallin} = \nlin_{m}^{\plin}$.
\end{lemma}

\begin{proof}
Using Theorems \ref{1plin-rstcl-complete} and \ref{cequallin-slrateq}, the well-known inclusion $\slrateq\subseteq\slrat$ yields the first containment  
$\cequallin\subseteq \plin$.
Next, we want to show that $\plin\subseteq \nlin_{m}^{\cequallin}$.
Let $A$ be any set in $\plin$. By Proposition \ref{1plin-1gaplin-complete}, choose a gap function $f\in\gaplin$ satisfying that $A=\{x\mid f(x)>0\}$. To simplify our proof, we assume that the empty string $\lambda$ is not in $A$. Let $N=(Q,\Sigma,\Gamma,q_0,q_{acc},q_{rej})$ be 
any linear-time counting 1TM that witnesses $f$. Without loss of generality, we can assume that (i) at each step, $N$ makes at most two nondeterministic choices and (ii) $f(x)\neq 0$ for all strings $x$ in $\Sigma^*$. Since $N$ runs in linear time, let $k$ be the minimal positive integer such that $\mathrm{Time}_{N}(x) < k|x|$ for any nonempty string $x$. It thus follows that $-2^{k|x|}< f(x)< 2^{k|x|}$.

For brevity, write $\Delta_k$ for the set $\{0,1\}^k$ and
assume a standard lexicographic order on the set $(\Delta_k)^*$ of strings. 
Let us define a reduction machine $M$ as follows. On input $x$, guess
a string, say $s$, over the alphabet $\Delta_k$ of length $|x|$ and 
produce $\track{x}{s}$ on the output tape by entering an accepting state. 
Note that there are exactly $2^{k|x|}$ nondeterministic branches. Note that the machine $M$ is meant to guess the value $f(x)+1$, if $f(x)>0$, and transfer this information to another machine $N'$. For each string $s\in(\Delta_k)^{|x|}$, let $l_s$ denote the positive integer $l$ for which $s$ is lexicographically the $l$th string in $(\Delta_k)^{|x|}$. Obviously, we have $1\leq l_s\leq 2^{k|x|}$. 
 
Next, we describe the counting 1TM $N'$. 
On input $\track{x}{s}$, $N'$ guesses a string $s'$ in 
$(\Delta_k)^{|x|}$ in the third
track. In case where $x=\lambda$, $N'$ rejects the input immediately and halt because $\lambda\not\in A$. Hereafter, we assume that $x\neq\lambda$. If $s<s'$, then $N'$ produces both an accepting path and a rejecting path, making no contribution to the gap function witnessed by $N'$. Consider the case where $s'=s$. In this case, $N'$ simulates $N$ on the input $x$. At length, when $s'<s$, $N'$ rejects the input immediately and halt. 
Note that $l_s = |\{s'\in (\Delta_k)^{|x|} \mid s'<s\}| + 1$. 
Let $g$ be the gap function induced by $N'$. 
For any nonempty string $x$ and any string $s\in (\Delta_k)^{|x|}$, we obtain $g(\track{x}{s}) = f(x) - (i_s-1)$. 

With the gap function $g$, we define a set $B=\{x\mid g(x)=0\}$, which is in $\cequallin$. For each string $x$, if $f(x)>0$, then $g(\track{x}{s})=0$ for
the $f(x)+1$st string $s$; otherwise, since $f(x)<0$, $g(\track{x}{s})$ is always negative for any choice $s\in (\Delta_k)^{|x|}$. This shows that $A$ is
many-one $1\mbox{-}\mathrm{NLIN}$-reducible to $B$ via $M$; namely, $A$ is in $\nlin_{m}^{B}$. Therefore, $A$ belongs to $\nlin_{m}^{\cequallin}$.

Finally, we want to show that $\nlin_{m}^{\cequallin} = \nlin_{m}^{\plin}$. The inclusion $\nlin^{\cequallin}_{m}\subseteq \nlin^{\plin}_{m}$ is obvious. Since $\plin\subseteq \nlin^{\cequallin}_{m}$,   
 $\nlin_{m}^{\plin}$ is contained 
in the complexity class $\nlin_{m}^{\nlin_{m}^{\cequallin}}$, 
which coincides with $\nlin_{m}^{\cequallin}$ by Proposition \ref{lemma:REG-closure}.
\end{proof}

The next proposition demonstrates two separation results concerning two complexity classes $\cequallin$ and $\plin$. Notationally, $\low_{m}\oneplin$ denotes the complexity class that is many-one low for $\oneplin$ (\ie $\low_{m}\oneplin =\{A\mid \oneplin_{m}^{A}\subseteq \oneplin\}$).

\begin{proposition}\label{plin-proper-inclusion}
\begin{enumerate}
\item $\oneplin\subsetneqq \onenlin_{m}^{\onecequallin}\cap \co\onenlin_{m}^{\onecequallin}$.
\vs{-2}
\item $\low_{m}\oneplin \subsetneqq \oneplin \subsetneqq \oneplin_{m}^{\oneplin}$.
\end{enumerate}
\end{proposition}

Proposition \ref{plin-proper-inclusion} is a consequence of 
the following key lemma regarding the complexity of 
the language $Center = \{ x1y \mid x,y\in\{0,1\}^*, |x|=|y| \}$. 

\begin{lemma}\label{center-in-class}
The language $Center$ belongs to $\onenlin_{m}^{\onecequallin}\cap\co\onenlin_{m}^{\onecequallin}$.
\end{lemma}

With help of this lemma, we can prove 
Proposition \ref{plin-proper-inclusion} easily. 

\begin{proofof}{Proposition \ref{plin-proper-inclusion}}
We have seen in Section \ref{sec:probabilistic} that $Center\not\in\mathrm{SL}_{rat}$. Assertion 1 follows directly from Lemma \ref{center-in-class}. The second part of Assertion 2  follows from Assertion 1 because $\onenlin_{m}^A\subseteq \oneplin_{m}^A$ for any set $A$. 
Next, we want to show the first part of Assertion 2. 
Let $A$ be any set in $\low_{m}\oneplin$, that is, 
$\oneplin_{m}^{A}\subseteq \oneplin$. Obviously, 
$A\leq^{\oneplin}_{m}A$, and thus $A\in\oneplin_{m}^{A}$.
This implies that $A\in\oneplin_{m}^{A}\subseteq \oneplin$.
Therefore, we obtain  $\low_{m}\oneplin\subseteq \oneplin$. 
The separation $\low_{m}\oneplin\neq \oneplin$ follows from the second part of Assertion 2 because $\low_{m}\oneplin = \oneplin$ implies $\oneplin = \oneplin_{m}^{\oneplin}$ by the definition of lowness.
\end{proofof}

To complete the proof of Proposition \ref{plin-proper-inclusion}, we still need to prove Lemma \ref{center-in-class}. The lemma can be proven by constructing many-one $\onenlin$-reductions from $Center$ to sets in $\onecequallin$.

\begin{proofof}{Lemma \ref{center-in-class}}
As a target set, we use the set $A=\{0^n\#1^n\mid n\in\nat \}$, where $\#$ is a special symbol not in $\{0,1\}$, which belongs to $\onecequallin$. 
First, we want to show that $Center \leq^{\nlin}_{m} A$, since this implies  that $Center\in \onenlin_{m}^{\onecequallin}$. Let us consider the following nondeterministic 1TM $N$ running in linear time. 
\begin{quote}
Let $x$ be any input. In Phase $1$, determine whether $|x|$ is odd and then return the head back to the start cell. If $|x|$ is even, output $x$ and halt immediately. Now, assume that $|x|$ is odd. In Phase $2$, choose nondeterministically either $0$ or $1$. If $1$ is chosen, go to Phase $3$; otherwise, overwrite $0$ in the scanning cell, move the head to the right, and then repeat Phase $2$. Whenever the head reaches the first blank symbol, return it to the start cell and halt. In Phase $3$, check whether the head is currently scanning $1$. If not, return the head back to the start cell and halt. Otherwise, change $1$ to $\#$ and then, by moving the head rightward, convert all the symbols on the right of $\#$ to $1$s. Finally, return the head to the start cell and halt.  
\end{quote}
We now prove that $Center\leq^{\onenlin}_{m}A$. Assuming that $x\in Center$, let $x=u1v$ for two strings $u$ and $v$ of the same length. Along a certain computation path, $N$ successfully converts $x$ to $0^{|u|}\#1^{|v|}$, which  belongs to $A$. On the contrary, assume that $x\not\in Center$. When $|x|$ is even, $N$ outputs $x$, which is obviously not in $A$. In case where $x$ is of the form $u0v$ with strings $u$ and $v$ of the same length, $N$ never outputs $0^{|u|}\#1^{|v|}$. Hence, $N$ many-one $\onenlin$-reduces $Center$ to $A$.

To show that $\overline{Center} \leq^{\nlin}_{m} A$, let us consider another nondeterministic 1TM $N'$ that behaves as follows.
\begin{quote}
On input $x$, check if $|x|$ is odd. If not, output $1\#1$ and halt. Otherwise, simulate Phase 2 of $N$'s algorithm. In Phase 3, check if the currently scanning cell has $1$. If so, return the head to the start cell and halt. Otherwise, overwrite $\#$ and convert the whole input $x$ into a string of the form $0^n\#1^m$ as an output. 
\end{quote}
A similar argument for $N$ can demonstrate that $N'$ many-one $\onenlin$-reduces $\overline{Center}$ to $A$. Hence, we conclude that $Center\in \co\onenlin_{m}^{\onecequallin}$. This completes the proof.
\end{proofof}

The complexity class that is many-one low for $\cequallin$,
denoted $\low_{m}\cequallin$, satisfies the inclusion relations
$\low_{m}\cequallin\subseteq\cequallin\cap\co\cequallin
\subsetneqq\cequallin$.
The first inclusion can be proven in a similar fashion to the proof of Corollary \ref{low-gaplin} and the second proper inclusion follows from Corollary \ref{cfl-pequallin}. Unlike Corollary \ref{low-gaplin}, it is open whether $\reg = \low_{m}\onecequallin$.

\begin{lemma}
$\low_{m}\cequallin\subseteq\cequallin\cap\co\cequallin
\subsetneqq\cequallin$.
\end{lemma}

Since $\onecequallin\subseteq \oneplin$, Proposition \ref{proposition:plin}  immediately yields the following separations: $\cfl\cap \reg/n \nsubseteq \cequallin$, $\cfl\nsubseteq \onecequallin\cap\reg/n$, 
and $\reg/n \nsubseteq \cfl\cup\cequallin$.
Other separations among three complexity classes 
$\cfl$, $\cequallin$, and $\reg/n$ are presented in the following  proposition. 

\begin{proposition}\label{lemma:non-uniform}
$\cequallin \cap\cfl \nsubseteq \reg/n$,  
$\cequallin\cap\reg/n \nsubseteq \cfl$, 
and
$\cequallin \nsubseteq \reg/n\cup\cfl$.
\end{proposition}

\begin{proof}
Let us consider the language 
$Equal = \{ w\in\{0,1\}^* \mid \#_0(w) = \#_1(w) \}$,
which stays outside of $\reg/n$. The first separation follows immediately since
$Equal$ belongs to $\cfl$ and $\cequallin$.
To prove the second claim,
recall the non-context-free language
$L_{3eq} = \{ a^nb^nc^n \mid n\in\nat \}$. 
We want to show that $L_{3eq}$ belongs to $\cequallin$. 
To see this, note that 
$L_1=\{a^nb^nc^m\mid m,n\in\nat\}$ and
$L_2=\{a^mb^nc^n\mid m,n\in\nat\}$ are in $\cequallin$. It is rather easy to show that the
intersection $L_1\cap L_2$ is also in $\cequallin$. Since 
$L_{3eq} = L_1\cap L_2$, $L_{3eq}$ belongs to $\onecequallin$. 

The second separation follows from the fact that $L_{3eq}\in\reg/n$.  
For the third separation, consider the non-context-free language
${\it 3}Equal$, given in Section \ref{sec:probabilistic}.
It is straightforward to show that ${\it 3}Equal$ belongs to $\cequallin$.
With a similar argument for the first separation, 
we can argue that ${\it 3}Equal$ cannot be in $\reg/n$.
\end{proof}

\section{Quantum Computation}\label{sec:quantum}

The notion of a quantum Turing machine was 
introduced by Deutsch \cite{Deu85} in the mid 1980s
and later reformulated by Bernstein and Vazirani
\cite{BV97} to model a quantum computation. 
Within our framework of 1TMs, 
we use a general model of
one-tape quantum Turing machines, which allow their tape 
heads to stay still \cite{Yam99a,Yam99b}.

A {\em (measure-once) one-tape quantum Turing machine} 
(abbreviated 1QTM) is similar to the
classical 1TM $(Q,\Sigma,\Gamma,\delta,q_0,q_{acc},q_{rej})$
except that its transition 
function $\delta$ is a map from 
$Q\times \Gamma$ to the vector space
$\complex^{Q\times \Gamma\times\{L,N,R\}}$. 
The {\em configuration space} of $M$ is 
the Hilbert space spanned by the set of all configurations of $M$ 
as the computational basis. Any element of this configuration space is called a {\em superposition of configurations}, which is a linear combination of configurations with complex coefficients (called {\em amplitudes}). A 1QTM $M$ is said to be {\em well-formed} if its time-evolution operator preserves the 
$\ell_2$-norm (\ie Euclidean
norm), where the {\em time-evolution
operator} for $M$ is the operator that maps a superposition of configurations to another superposition of the
configurations resulting by an application of the quantum
transition function $\delta$ of $M$. 
For any subset $K$ of $\complex$, a 1QTM is said to have {\em
$K$-amplitudes} if all amplitudes in $\delta$ 
are drawn from $K$.
By ignoring its nonzero transition amplitudes, $\delta$ can be viewed 
as a nondeterministic transition function. For clarity, we use the notation $\hat{\delta}$ to express this nondeterministic transition function. Similar to the classical case, a {\em (classical) computation path} of a 1QTM is defined as a series of configurations, each of which is obtained from its previous configuration  by an application of $\hat{\delta}$. These classical computation paths form a {\em classical computation tree}. Any quantum computation can be viewed as its corresponding classical computation tree in which each edge is weighted by its associated nonzero amplitude. 

Unlike classical Turing machines, there is a subtle but arguable 
issue concerning the definition of the halting condition of a 1QTM. In accordance with the classical halting condition, we define the {\em running time} of a 1QTM $M$ on input $x$ as the minimal nonnegative integer $t$ such that, in the classical computation tree $T$ representing the quantum computation of $M$ on the input $x$, all configurations in $T$ become halting configurations at time $t$ {\em for the first time}. If such a $t$ exists, we say that {\em $M$ halts\footnote{This definition comes from Bernstein and Vazirani \cite{BV97}, who defined a quantum Turing machine to ``halt'' at time $t$ if the superposition of configurations at time $t$ consists only of halting configurations and, at time less than $t$, the superposition contains no halting configuration. See also \cite{Yam99a,Yam99b} for more discussions.} at time $t$}. This halting condition makes us view  time-bounded 1QTMs as classical ``synchronous'' machines. 
A time-bounded 1QTM $M$ is said to be {\em well-behaved} if, when $M$ halts,
the tape head halts in the same cell 
(not necessarily the start cell) in all halting configurations of the classical computation tree representing the quantum computation of $M$. Moreover, $M$ is {\em stationary} if it is well-behaved and the head always halts in the start cell. 

The {\em acceptance probability} of a 1QTM $M$ on input $x$, denoted $p_{M}(x)$, is 
the sum of all the squared magnitudes of  
accepting configurations 
(\ie configurations with the internal state $q_{acc}$) in
any superposition generated at the time when $M$ halts on the input $x$.
Let $K$ be any
nonempty subset of $\complex$. We introduce the {\em one-tape bounded-error quantum linear-time} class $\bqlin_{K}$ as the collection of all languages $L$ that satisfy the following condition: there exist a linear-time well-formed 
stationary 1QTM $M$ with $K$-amplitudes and an error bound
$\varepsilon>0$ such that, for every string $x$, 
(i) if $x\in L$, then
$p_{M}(x) \geq 1/2+\varepsilon$ and
(ii) if $x\notin L$, then $p_{M}(x)\leq 1/2-\varepsilon$.

It is important to note that our linear-time 1QTMs may not simulate linear-time 2-way quantum finite automata given in \cite{KW97} mainly because of the synchronous condition of our 1QTMs. On the contrary, the synchronous condition enables us to prove in Lemma \ref{qtm-gaplin} a strong connection between 1QTMs and $\gaplin$.

We prove a key lemma, which shows how to compute the acceptance probability of a 1QTM with $\rational$-amplitudes. The lemma has a similar flavor to Theorem 3(4) in \cite{Yam99b} (see also \cite{FR97}). In the following proof, we use the folding machine obtained from a given 1QTM. 

\begin{lemma}\label{qtm-gaplin}
Let $M$ be any well-formed stationary 1QTM with $\rational$-amplitudes. If $M$ always halts, then there exist a constant $d\in\nat^{+}$ and a function $f$ in $\gaplin$ such that $p_{M}(x)=f(x)\cdot d^{-\mathrm{Time}_{M}(x)}$ for every string $x$.
\end{lemma}

\begin{proof}
Given a 1QTM $M$, Since the construction of a folding machine, given in Section \ref{partial-function}, is applicable to any 1QTM, 
we can work on $M$'s folding machine $N=(Q,\Sigma,\Gamma,\delta,q_0,q_{acc},q_{rej})$, which simulates $M$ in $N$'s tape using only the input area.
Notice that $N$ may violate unitarity and no longer be well-formed.
Since $N$ uses rational amplitudes, we can choose the minimal integer $c\in\nat^{+}$ satisfying that every amplitude of $N$ has the form $r/c$, where $r$ is a certain integer. Fix $x$ arbitrarily and let $y$ be any (classical) computation path of $N$ on input $x$. When $N$ halts, an accepting configuration of $N$ depends only on the tape content because $N$'s internal state and its head position in the accepting configuration are predetermined. It thus suffices to consider a final tape content of $N$. Let $z$ be any final tape content of $N$. Note that $|z|=|x|$ since $N$ rewrites only the contents of cells in the input area. We denote by $amp_{N}(x,y,z)$ the amplitude associated with accepting computation path $y$ of $N$ on input $x$ leading to the final tape content $z$. Since $N$ is synchronous, the  value $amp_{N}(x,y,z)\cdot c^{\mathrm{Time}_{M}(x)}$ is always an integer. 

Now, we define the function $f_{+}$ as $f_{+}(\track{x}{z})= c^{\mathrm{Time}_{M}(x)}\cdot\sum_{y}amp_{N}(x,y,z)$, where the sum $\sum$ is taken over all accepting computation paths $y$ of $N$ on input $x$ that leads to the final tape content $z$ with {\em positive} amplitude. We want to show that $f\in \flin$. For our purpose, we first translate the 1QTM $N$ into a  classical nondeterministic 1TM $\hat{N}$ in such a way that, whenever $N$ makes a transition with a transition amplitude of the form $m/c$ for certain integers $m,c$ with $c>0$, $\hat{N}$ produces exactly $|m|$ nondeterministic branches. As a result, for each of such $y$'s, we can generate exactly $amp_{N}(x,y,z)\cdot c^{\mathrm{Time}_{M}(x)}$ branches leading to certain  accepting configurations. To determine the sign of the amplitude associated with each computation path of $N$, we further prepare two sets of internal states for $\hat{N}$ and move one set to another whenever the amplitude sign changes by an application of $\delta$. The resulting machine witnesses that 
$f_{+}$ is in $\sharplin$. Similarly, we define $f_{-}(\track{x}{z}) = c^{\mathrm{Time}_{M}(x)}\cdot\sum'_{y}amp_{N}(x,y,z)$, where the sum $\sum'$ is taken over all accepting computation paths of $N$ on input $x$ that leads to the final tape content $z$ with {\em negative} amplitude. We also conclude that  $f_{-}\in\sharplin$.

Recall that the acceptance probability $p_{M}(x)$ is the sum of $(\sum_{y}amp_{N}(x,y,z) - \sum'_{y}amp_{N}(x,y,z))^2$ over all possible final tape contents $z$ of $N$; in other words, 
$c^{2\mathrm{Time}_{M}(x)}\cdot p_{M}(x) = \sum_{z\in\Gamma^{|x|}}(f_{+}(\track{x}{z}) - f_{-}(\track{x}{z}))^2$. {}From the closure property of $\gaplin$ (Lemma \ref{gaplin-closure}), the function appearing in the right-hand side of the last equation clearly belongs to $\gaplin$. For the desired constant $d$, we set $d=c^2$. 
This completes the proof.
\end{proof}

A simple application of Lemma \ref{qtm-gaplin} shows the following proposition. 

\begin{proposition}\label{r1TM-is-1QTM}
  $\reg\subseteq
  \bqlin_{\rational} \subseteq \plin$.
\end{proposition}

\begin{proof}
Note that every (deterministic) 
reversible 1TM can be viewed as a
well-formed 1QTM with $\rational$-amplitudes,   
which produces no computational error.  
{}From this, it follows that $\rdlin\subseteq \bqlin_{\rational}$.
Proposition \ref{trev} therefore implies that 
$\reg\subseteq \bqlin_{\rational}$. 

We want to show the second inclusion. Let $L$ be any language in $\bqlin_{\rational}$ and choose a linear-time well-formed 1QTM $M$ that recognizes $L$ with bounded-error probability. Moreover, $M$ uses only rational amplitudes. By amplifying the success probability (by, \eg  a majority vote technique), we can assume without loss of generality that, for every $x$, either $p_{M}(x)\geq 2/3$ or $p_{M}(x)\leq 1/3$. By Lemma \ref{qtm-gaplin}, we find a constant $d\in\nat^{+}$ and a function $f\in\gaplin$ such that 
$f(x)=p_{M}(x)\cdot d^{\mathrm{Time}_{M}(x)}$ for every input $x$. Now, we define $g(x)= d^{\mathrm{Time}_{M}(x)}$ for each string $x$. It thus follows that $x\in L$ implies $3f(x)>2g(x)$ and that $x\not\in L$ implies $3f(x)<g(x)$. To complete the proof, we need to define $h(x)=3f(x)-2g(x)$, which is also in $\gaplin$ by Lemma \ref{gaplin-closure}. This $h$ satisfies $L=\{x\mid h(x)>0\}$. Hence, $L$ is in $\plin$. 
\end{proof}

A variant of quantum Turing machine, 
so-called a ``nondeterministic'' quantum 
Turing machine, which is
considered as a quantum analogue of a 
nondeterministic Turing machine, was
introduced by Adleman et al. \cite{ADH97}. 
Let $K$ be any nonempty subset of $\complex$.  A
language $L$ is in $\nqlin_{K}$ if there 
exist a linear-time
well-formed stationary 1QTM $M$ 
with $K$-amplitudes
such that, for every $x$, $x\in L$ if and only if $M$
accepts input $x$ with positive probability.
 
We show that $\nqlin_{\rational}$ can be precisely
characterized by linear-time counting 1TMs. This result can be compared with a polynomial-time case of 
$\nqp_{\complex}=\co\cequalp$ \cite{YY99}.

\begin{proposition}\label{nqp-character}
$\nqlin_{\{0,\pm3/4,\pm4/5,\pm1\}} = \nqlin_{\rational} = 
\co\cequallin$.
\end{proposition}

Proposition \ref{nqp-character} is obtained by combining two lemmas: Lemmas \ref{nqlin-slrateq} and \ref{slrateq-nqlin}.

\begin{lemma}\label{nqlin-slrateq}
$\nqlin_{\rational}\subseteq \co\cequallin$.
\end{lemma}

\begin{proof}
Let $L$ be any language in $\nqlin_{\rational}$. Choose a linear-time well-formed stationary 1QTM $M$ satisfying that $L=\{x\mid p_{M}(x)>0\}$. Applying Lemma \ref{qtm-gaplin}, we obtain a constant $d\in\nat^{+}$ and a function $f$ in $\gaplin$ such that $f(x)=p_{M}(x)\cdot d^{\mathrm{Time}_{M}(x)}$ for any string $x$. It immediately follows that, for every $x$, $x\in L$ if and only if $f(x)\neq 0$. Therefore, $L$ belongs to the complement of $\cequallin$.
\end{proof}

Finally, we prove the remaining inclusion
$\co\cequallin \subseteq\nqlin_{\{0,\pm3/5,\pm4/5,\pm1\}}$. {}From the fact $\cequallin=\slrateq$, it suffices to show that $\co\slrateq \subseteq \nqlin_{\{0,\pm3/4,\pm4/5,\pm1\}}$. In the proof of Lemma \ref{slrateq-nqlin}, we use the following two unitary 
transformations $U$ and $V$ acting on $\{\ket{s}\mid s\in [0,3]_{\integer}\}$. Let $U\ket{0}=\frac{3}{5}\ket{0}+\frac{4}{5}\ket{1}$, $U\ket{1}=-\frac{4}{5}\ket{0}+\frac{3}{5}\ket{1}$, and $U\ket{s}=\ket{s}$ for $s\in\{2,3\}$. Let $V\ket{s}=\frac{(1-s)4}{5}\ket{s}+\frac{3}{5}\ket{(s+2)\!\mod 4}$ for $s\in\{0,2\}$ and $V\ket{s}=\frac{3}{5}\ket{s}+ \frac{(s-2)4}{5}\ket{(s+2)\!\mod 4}$ for $s\in\{1,3\}$.

\begin{lemma}\label{slrateq-nqlin}
$\co\slrateq\subseteq \nqlin_{\{0,\pm3/4,\pm4/5,\pm1\}}$.
\end{lemma}

\begin{proof}
Let $L$ be any set in $\co\slrateq$. There exists a rational 1PFA
$N=(S, \Sigma,\pi,\{T(\sigma)\,|\,\sigma\in\Sigma\}, \eta)$ such that  $\overline{L} = L^{=}(N,\epsilon)$  for
a certain rational cut point $\epsilon$. Similar to the proof of 
Lemma \ref{rstcl-subset-1synplin}, we can assume that (i) $L=\{x\in\Sigma^*\mid p_N(x)\neq1/2\}$, (ii) $S=[1,\ell]_{\integer}$ for a certain number $\ell\in\nat^{+}$, (iii) one component of $\pi$ is 1, and
(iv) there is a positive integer $m$ satisfying the following property: for any $\sigma\in\Sigma$ and any $i,j\in S$,
$T(\sigma)_{i,j}$ is of the form $r_{i,j}(\sigma)/2^m$ for a certain number $r_{i,j}(\sigma)\in\nat$. Let $F$ be the set of all final states of $N$.

Hereafter, we wish to construct
a linear-time well-formed stationary 1QTM $M$
with $\{0,\pm3/5,\pm4/5,\pm1\}$-amplitudes and show that, 
for any 
nonempty string $x$, $p_{M}(x)>0$ if and only if $p_N(x)\neq 1/2$. {}From this, we can conclude that $L$ belongs to $\nqlin_{\rational}$. 
Let $x=\sigma_1\dots\sigma_n$ be any string, where each symbol 
$\sigma_j$ is in $\Sigma$ and $n\ge 0$. 
Let $\Delta=\{0,1\}^m$ be our new alphabet. Assuming a linear order on $\Delta$, for each symbol $k\in \Delta$, we define $l_k$ to be the number satisfying that $k$ is the $l_k+1$st symbol in $\Delta$. Note that $0\leq l_k<2^m$ for any $k\in\Delta$.

\ms

1) Initially, $M$ is in the initial state $q_0$, scanning
the start cell (indexed $0$). If $x=\lambda$, then $M$ immediately accepts or rejects the input if $\lambda\in L$ or $\lambda\not\in L$, respectively. In the rest of the description of $M$, we assume that $|x|\geq1$. 
In this preprocessing phase, $M$ replaces each input symbol $\sigma$ by its corresponding new symbol $\track{\sigma}{0^m}$ by moving its tape head rightward. In the end, $M$ returns the tape head to the start cell. In the subsequent description of $M$, we pay our attention to the content of the cells indexed between $0$ and $n$. 

2) The machine $M$ simulates a series of ``coin flips'' of $N$ by generating a certain superposition of configurations. By moving the tape head rightward again, 
$M$ applies the transformation $U^{\otimes m}$ to the symbol $0^m$ given
in $\track{\sigma}{0^m}$: $U^{\otimes m}\ket{0^m} = 
\sum_{k\in \Delta}\left(\frac{3}{5}\right)^{\#_0(k)}
\left(\frac{4}{5}\right)^{\#_1(k)} \ket{k}$, where $\#_i(k)$ denotes the number of $i$'s in $k$ when $k$ is viewed as an $m$-bit string. 
When $M$ reaches the first blank symbol (in the $n$th cell),
it returns the tape head to the start cell. On each (classical) computation path, the tape content must be in the form   
$\track{x}{\vec{k}} = \track{\sigma_1}{k_1}\track{\sigma_2}{k_2}\cdots\track{\sigma_n}{k_n}$, where $\vec{k} = k_1k_2\cdots k_n\in \Delta^n$. 

3) Assume that $N$'s initial state is $0$. Let $p_0$ be a new internal state of $M$ associated with $N$'s. Now, we make $M$ simulate each step of $N$ in such a way that, when $N$ makes a transition from an internal state $a$ to another state $b$ with transition probability $r_{a,b}$ for an input symbol $\sigma$, $M$ generates exactly $r_{a,b}$ (classical) computation paths. This can be done with new internal states $p_a$ and $p_b$ and by applying 
the following transition rule: for every symbol $k\in\Delta$, 
$\delta(p_{a},\track{\sigma}{k})
=\ket{p_{1}}\ket{\track{\sigma}{k}}\ket{R}$
if $0\le 
l_k< r_{a,1}(\sigma)$ and $\delta(p_{a},\track{\sigma}{k})
=\ket{p_{b}}\ket{\track{\sigma}{k}}\ket{R}$
if $b>1$ and $\sum_{i=1}^{b-1}r_{a,i}(\sigma)\le 
l_k<\sum_{i=1}^{b}r_{a,i}(\sigma)$. 

4) After reaching the $n$th cell in internal state $p_{a}$, $M$ writes down the outcome $0$ or $1$ of the $N$ depending on $a\in F$ or $a\not\in F$, respectively, and then enters a new internal state $s_a$. 
When $\track{x}{\vec{k}}$ is produced in phase 2, 
let $r_{\vec{k}}$ denote this outcome written in the $n$th cell. Note that
$p_N(x)= |\{\vec{k}\in\Delta^{n}\mid r_{\vec{k}}=0\}|\cdot2^{-mn}$.

5) In this phase, $M$ first reverses phase 3. 
This brings the tape head back to the start cell and the internal state to $p_0$. By moving the head rightward again,
$M$ also applies $U^{\otimes m}$ to each symbol $k$
in $\track{\sigma}{k}$, just as in phase 2, collapsing at most $2^m$ branches to each configuration containing tape content $\track{x}{\vec{k}}\,r$, where $\vec{k}\in\Delta^n$ and $r\in\{0,1\}$.
In particular, we obtain the configuration with $\track{x}{1^{mn}}\,r = \track{\sigma_1}{1^m}\track{\sigma_2}{1^m}\cdots \track{\sigma_n}{1^m}\,r$ 
with  amplitude $amp(r) = \left(\frac{12}{25}\right)^{mn} |\{\vec{k}\in\Delta^{n}\mid r_{\vec{k}}= r\}|$.

6) We need to make the accepting paths and rejecting paths of $N$ interfere to each other. This is done by applying the unitary transformation $W=UV$ to the $n$th cell since $W$ maps $\ket{0}$ and $\ket{1}$ to $\ket{0}$ with amplitude $12/25$ and $-12/25$, respectively. 

7) Finally, $N$ checks if the cells indexed between $0$ and $n$ consists of $\track{x}{1^{mn}}\,0$. If so, $M$ enters $q_{acc}$; otherwise, $M$ enters $q_{rej}$. This phase can be done in a reversible fashion. This completes the description of $M$.

\ms

For any nonempty string $x$, a simple calculation shows that 
the acceptance probability $p_M(x)$ of $M$ is $p_{M}(x)=
  \left(\frac{12}{25}\right)^2\left( amp(0) - amp(1) \right)^2$, which equals $\left(\frac{24}{25}\right)^{2mn+2}(p_N(x)-1/2)^2$. It therefore follows that $p_{M}(x)>0$ if and only if $p_{N}(x)\neq 1/2$, as requested. 
\end{proof}

\section{Epilogue}

By exploring the close relationships to 
automata theory, we have studied the computational complexity 
of one-tape linear-time Turing machines of 
various machine types.
Since these machines are relatively weak in power,
we have proven the collapses and separations of 
several complexity classes without any unproven assumptions. Hennie's crossing sequence arguments and various simulation techniques are proven to be viable tools throughout this paper. Nonetheless, we have left numerous questions unsolved. Challenging these questions may bring in new proof techniques.

For further research on the theory of
one-tape linear-time Turing machines, we suggest five important 
future directions of the research.
\begin{itemize}
\item[i)] The model of Turing machines has significantly 
evolved over the past four decades. We have shown in this paper that different machine types can alter the power of computation. There are many more machine types that we have not yet discussed in this paper. Other types of Turing machines include metric Turing machines, bottleneck Turing machines, and interactive Turing machines (see, \eg \cite{DK00,HO02,Kre88}). We need to explore the computational power of such models and study the properties of complexity classes induced in terms of these models.
\vs{-2}\item[ii)] 
Despite the ability to 
alter the tape content, we have shown that many Turing machines  
working as language recognizers
cannot be more powerful than their associated finite state 
automata. To study the power of Turing machines, we need to explore their special ability to compute ``functions'' instead. In the past, such functions have been studied extensively in terms of {\em search problems}, {\em optimization problems}, and {\em approximation problems}. The study of these functions may present different perspectives to our understandings of one-tape computation. 
\vs{-2}
\item[iii)]
It is natural to ask what is the most complex language existing in a given complexity class. The theory of NP-completeness, for instance, sheds light on this question using various polynomial-time reductions. On the contrary, most one-tape linear-time complexity
classes that we have studied in this paper are unlikely to possess ``complete'' problems via many-one $\dlin$-reductions. Is there any ``weak'' reducibility that highlights the relative complexity of languages? 
\vs{-2}
\item[iv)]
We have considered advised computations; however, the role of advice has not been fully studied in this paper. It is important to investigate how much extra power advice can give to an underlying computation. Moreover, advised computations are often characterized by non-uniform computations. We also need to study the non-uniformity of one-tape linear-time computations in connection to advice.
\vs{-2}
\item[v)]
Relativization has had a great success in the polynomial-time complexity theory. Throughout this paper, we have studied only many-one relativization since many-one relativization is of the simplest form. The investigation of other types of meaningful relativization is also necessary for one-tape linear-time complexity classes.  
\end{itemize}
We hope that the further study of the above structural complexity issues on resource-bounded computations will lead to the better understandings of the effect of bounded resources of Turing machines

\bs
\paragraph{Acknowledgments.}
The first author is grateful to
the Japan Science and Technology Corporation
and the 21st Century COE Security 
Program of Chuo University for financial support.
He also thanks Masahiro Hachimori for 
valuable suggestions on
probabilistic computations. The second author 
thanks Harumichi Nishimura and Raymond H.
Putra for fruitful discussions on Turing machines. The authors gracefully appreciate critical comments of anonymous referees. 

\bibliographystyle{alpha}

\section*{Appendix}

We show the proofs of Lemmas 
\ref{nlogn-bound} and \ref{nbound-regular} for completeness.

\begin{proofof}{Lemma \ref{nlogn-bound}}
Let $M$ be given as in the lemma and let $q$ be the number of internal states
of $M$. By its definition, $M$ has at least three states (that is,
$q\geq3$). Choose a number $n_0\in\nat$ such that $T(n)>0$ for all numbers $n\geq n_0$. We define the function $f$ from $\{n\in\nat\mid n\geq n_0\}$ to $\real^{\geq0}$
by $f(n) = \sqrt{n\log n/T(n)}$. Since $T(n)=o(n\log n)$, it follows that 
$\lim_{n\rightarrow\infty}f(n)=\infty$. Choose the smallest number
$c\in\nat$ such that, for every $n\ge 2$,
\[
\frac{3\left(q^{\frac{\log n}{f(n)}+1}-1\right)}{q-1}\le
n\left(1-\frac{1}{f(n)}\right)+ \frac{c\cdot f(n)}{\log n} + 1.
\]
Such $c$ exists because $q^{\frac{\log n}{f(n)}+1}=o(n)$.

Assume to the contrary that there exist a crossing sequence $\gamma$
of length longer than $c$ and an input $x$ ($|x|\geq 2$) such that
$\gamma$ is a crossing sequence at a certain critical-boundary $b$ of $x$
along a certain (accepting or rejecting) computation path $s$ of $M$ on $x$. Such a
crossing sequence $\gamma$ is called {\em long}, and other crossing
sequences are called {\em short}.

Let $x_0$ denote lexicographically the first input string that has a long
crossing sequence. Let $n_0=|x_0|$. Let $s_0$ be the shortest
computation path of $M$ on the input $x_0$ that generates a long
crossing sequence. Note that $|s_0|\leq T(|x_0|)$ by our
assumption. Moreover, let $b_0$ be the leftmost intercell boundary in
the tape that corresponds to a certain long crossing sequence, say
$\gamma_0$, along the computation path $s_0$.

Let us consider all critical boundaries of $x_0$ along the path 
$s_0$ whose crossing sequences are of lengths at most 
$\log n_0/f(n_0)$.  Let $h$ be the
number of all such critical boundaries.  Since the total computation
steps along the path $s_0$ is equal to the sum of the lengths of any crossing
sequences at intercell boundaries, we have
$T(n_0)>c+(n_0 +1 - h)\,\frac{\log n_0}{f(n_0)}$.
The inequality comes from the assumption that the length of $\gamma_0$
is longer than $c$.  Thus, we have
\begin{equation*}
  \frac{h}{3} > \frac{1}{3}\left( n_0 +1 -\frac{n_0}{f(n_0)} + 
  \frac{c\cdot f(n_0)}{\log n_0}\right) \geq
  \frac{q^{\frac{\log n_0}{f(n_0)}+1}-1}{q-1}\ge
  \sum_{i=0}^{\lfloor\log n_0/f(n_0)\rfloor} q^i,
\end{equation*}
which is at least the number of all crossing sequences 
of lengths at most 
$\log n_0/f(n_0)$.  Hence, there exist at least four
distinct critical boundaries $b_1,b_2,b_3,b_4$ that have an identical
crossing sequence in the path $s_0$.  Clearly, at least two of them (say
$b_1$ and $b_2$) are on the same side of $b_0$.

Now, we delete the region between $b_1$ and $b_2$ from the tape.  Let
$x_0'$ be the input string obtained from $x_0$ by this deletion.
Clearly, $|x_0'|<|x_0|$.  Moreover, the new path obtained from $s_0$
by this deletion is a valid computation path of M on the input $x_0'$
and still has a crossing sequence whose length is greater than $c$.  This
contradicts the minimality of $x_0$. Therefore, the lemma holds.
\end{proofof}

\begin{proofof}{Lemma \ref{nbound-regular}}
Let $n$ be any number in $\nat$.  For each string $x\in\Sigma^{\leq n}$ and
each crossing sequence $v\in S_n$, we say that $x$ {\em $n$-supports} $v$ if there
exists a string $z$ such that (i) $|xz|\leq n$, (ii) $xz\in L$, and
(iii) $v$ is the crossing sequence at the intercell boundary between
$x$ and $z$ along a certain {\em accepting} computation path of $M$ on the 
input $xz$. Now, let $\Supp{n}{x}=\{v\in
S_n \mid \mbox{ $x$ $n$-supports $v$ }\}$.

We want to show that, for any three strings $x,y,z\in\Sigma^*$, 
if $|xz|\leq n$, $|yz|\leq n$, $xz\in L$,
and $\Supp{n}{x}=\Supp{n}{y}$, then $yz\in L$.
This is shown as follows. Assume that $xz\in L$. Let $v$ be any crossing
sequence between $x$ and $z$ along a certain accepting computation path of
$M$ on the input $xz$. Clearly, we have $v\in \Supp{n}{x}$. Since
$\Supp{n}{x}=\Supp{n}{y}$, there exists a string $z'$ such that $v$ is
a crossing sequence between $y$ and $z'$ along an accepting
computation path of $M$ on $yz'$. Assume that the tape head halts in
the left region of $v$. Consider any computation of $M$ on the input
$yz$. By the nature of crossing sequences, $yz$ has an accepting
computation. Thus, we conclude that $yz\in L$. 
Similarly, we obtain the same conclusion
in the case where the tape head halts in the right region of $v$.

Note that $N_{L}(n)$ should be at most the number of distinct sets 
$\Supp{n}{x}$ over all strings $x\in\Sigma^{\leq n}$. Therefore, $N_{L}(n)$
is upper-bounded by $2^{|S_n|}$.
\end{proofof}

\end{document}